\documentclass[a4paper,pre,reprint,superscriptaddress,amsmath,amssymb,floatfix]{revtex4-2}

\usepackage[charter]{mathdesign}
\DeclareSymbolFont{usualmathcal}{OMS}{cmsy}{m}{n}
\DeclareSymbolFontAlphabet{\mathcal}{usualmathcal}
\usepackage{graphicx}% Include figure files
\usepackage{bm}% bold math
\usepackage{natbib}
\usepackage[dvipsnames]{xcolor}
\usepackage{hyperref}
\hypersetup{  colorlinks,
              linktoc         = page, % section, page, all, none
              linkcolor       = {orange!60!black}, % footnotes+ref's
              urlcolor        = {magenta!80!black}, % reference links
              citecolor       = {magenta!80!black}, %~\cite's
              anchorcolor     = {yellow}
}
\usepackage[top=2cm,bottom=2.5cm,left=1.7cm,right=1.5cm]{geometry}

\newcommand{\panel}[1]{{\fontfamily{phv}\selectfont\textbf{#1}}}

\renewcommand{\selectlanguage}[1]{}

\begin{document}

\title{Digital holographic imaging for free surfaces of superfluid helium}

\author{Vitor S. Barroso}
\email{vitor.barroso.s@outlook.com}
\affiliation{School of Mathematical Sciences, University of Nottingham, University Park, Nottingham, NG7 2RD, UK}

\author{Patrik \v{S}van\v{c}ara}
\email{patrik.svancara@neel.cnrs.fr}
\altaffiliation[Present address: ]{Institut N\'{e}el, CNRS-UGA, 25 avenue des Martyrs, 38042 Grenoble, France}
\affiliation{School of Mathematical Sciences, University of Nottingham, University Park, Nottingham, NG7 2RD, UK}

\author{Chris Goodwin}
\affiliation{School of Mathematical Sciences, University of Nottingham, University Park, Nottingham, NG7 2RD, UK}

\author{Sreelekshmi C. Ajithkumar}
\altaffiliation[Present address: ]{Department of Physics, University of Strathclyde, Glasgow, G4 0NG, United Kingdom}
\affiliation{School of Physics and Astronomy, University of Nottingham, University Park, Nottingham, NG7 2RD, UK}

\author{Ilaria Dimina}
\affiliation{School of Mathematical Sciences, University of Nottingham, University Park, Nottingham, NG7 2RD, UK}

\author{Silvia Schiattarella}
\affiliation{School of Physics and Astronomy, University of Nottingham, University Park, Nottingham, NG7 2RD, UK}

\author{Pietro Smaniotto}
\affiliation{School of Mathematical Sciences, University of Nottingham, University Park, Nottingham, NG7 2RD, UK}

\author{Leonardo Solidoro}
\affiliation{School of Mathematical Sciences, University of Nottingham, University Park, Nottingham, NG7 2RD, UK}

\author{Marion Cromb}
\affiliation{School of Mathematical Sciences, University of Nottingham, University Park, Nottingham, NG7 2RD, UK}

\author{Radivoje Prizia}
\altaffiliation[Present address: ]{Laboratoire Kastler-Brossel, Sorbonne Universit\'{e}, ENS-Universit\'{e} PSL, CNRS, Coll\`{e}ge de France, 4 place Jussieu, 75005 Paris, France}
\affiliation{School of Physics and Astronomy, University of Nottingham, University Park, Nottingham, NG7 2RD, UK}

\author{Anthony J. Kent}
\affiliation{School of Physics and Astronomy, University of Nottingham, University Park, Nottingham, NG7 2RD, UK}

\author{Silke Weinfurtner}
\affiliation{School of Mathematical Sciences, University of Nottingham, University Park, Nottingham, NG7 2RD, UK}
\affiliation{Department of Physics and Astronomy, University of Manchester, Manchester, M13 9PL, UK}
\affiliation{Photon Science Institute, Alan Turing Building, University of Manchester, Manchester, M13 9PY, UK}

\begin{abstract}
Visualising the free surface of superfluid helium offers a rare opportunity to explore wave dynamics in the limit of vanishing viscosity. Such measurements are nonetheless challenging due to helium's low refractive index contrast, restricted optical access to the cryogenic setups required to maintain helium in its superfluid phase, and mechanical vibrations from the various cooling stages. Overcoming these limitations will enable quantitative studies of surface-wave dynamics with applications in fluid mechanics, quantum simulation, and quantum optomechanics. Here we report an implementation of off-axis digital holography for full-field imaging of the free surface of superfluid \textsuperscript{4}He. We perform non-contact measurements of nanometre- to micrometre-scale interface fluctuations in two cryogenic systems: a traditional helium bath cryostat and a cryogen-free refrigerator. We employ machine-learning-based analysis to isolate noise-driven normal modes and their spatial structure in both systems. This enables reconstruction of the dispersion relation for gravity-capillary waves in macroscopic samples and, for thick films, determination of the film thickness from the measured dispersion, providing a quantitative benchmark for our approach. These proof-of-concept experiments show that digital holography is a powerful and versatile tool for high-resolution, minimally invasive studies of superfluid surfaces, with strong potential for integration into diverse experimental platforms.
\end{abstract}

\maketitle

\section{\label{sec:intro}Introduction}

Wave dynamics on free fluid surfaces exemplify a broader class of non-equilibrium phenomena central to modern many-body physics. Here, we focus on waves propagating on the free surface of superfluid \textsuperscript{4}He, whose exceptionally low viscosity allows the study of nonlinear wave dynamics in compact laboratory settings~\cite{Kolmakov2009,kolmakov2014wave}. However, high-resolution imaging of the superfluid interface is challenging and requires adapting optical methods to cryogenic environments. Several such approaches have been developed to date. For instance, investigations of vertically shaken superfluid samples have utilised laser beam reflection from the interface, leading to the observation of a cascade of Faraday waves~\cite{Abdurakhimov2010-vt,Mezhov-Deglin2019-hs}. Although this technique offers excellent and tunable frequency resolution, it is limited to probing a single spot on the interface. In contrast, interferometry has been used to image surface deformations in rotating superfluid helium~\cite{Marston1977-lx}, to measure the morphology and growth of solid \textsuperscript{3}He crystals~\cite{tsepelin2002morphology, todoshchenko2007growth}, and has also been proposed as a means of detecting small, localised depressions of the free surface caused by individual quantum vortices, naturally occurring in superfluid helium~\cite{Seddon2003-zj}. However, the predicted, approximately $7$-nm deep quantum dimples~\cite{Vittoratos1973-xm} are yet to be confirmed experimentally~\cite{Seddon2005-es}, owing to the mechanical noise present in cryogenic systems that excites surface waves of much larger amplitude. 

A significant advancement in accessing the interface between helium vapour and superfluid was brought by the implementation of synthetic Schlieren imaging~\cite{Moisy2009-qm}. The approach relies on a regular two-dimensional background pattern that is imaged through the superfluid's free surface by a high-speed digital camera. Despite helium's low refractive index ($n = 1.028$~\cite{donnelly1998}), waves propagating on the interface induce detectable and deterministic deformations of the pattern. These deformations are analysed in the Fourier domain~\cite{Wildeman2018-cn} and compared against a still pattern, revealing the surface topography as a function of both space and time. This method was used to resolve surface waves with micrometric amplitudes in superfluid helium~\cite{Svancara2024-ac}, enabling the study of their interactions with a background vortex flow~\cite{Svancara2024-ac,Smaniotto2025-bl}. Despite its success in reconstructing superfluid free surfaces, synthetic Schlieren imaging only provides the topography indirectly, as it is fundamentally limited to measuring surface slopes rather than absolute heights. Additionally, its spatial resolution is constrained by the pattern's design and imaging contrast.

Off-axis digital holography (DH)~\cite{yaroslavsky2013digital,picart2013digital,schnars2014digital} offers a pathway towards time- and space-resolved measurements of interface deformations at the nanometre scale. This technique enables quantitative retrieval of phase shifts from holograms formed by the interference of two beams: a reference beam propagating through free space and a probe (or object) beam transmitted through the liquid sample. Surface waves induce variations in the optical path of the probe beam, producing local phase shifts that are recorded in the digital hologram. Reconstructing these phase shifts directly yields the topography of the evolving surface, obtained by scaling them with a constant factor proportional to the refractive index contrast $\Delta n$ between the sample and free space~\cite{bhaduri2014diffraction,S_Barroso2023-zh}. This very principle has found numerous applications across multiple fields~\cite{huang2024quantitative}, including biomedicine~\cite{park2018quantitative}, microscopy~\cite{yu2014review}, microfluidics~\cite{wu2012optical}, industrial metrology~\cite{kumar2023emerging} and even the imaging of quantum gases~\cite{smits2020imaging}.

\begin{figure*}[tbp!]
    \centering
    \includegraphics[scale=1]{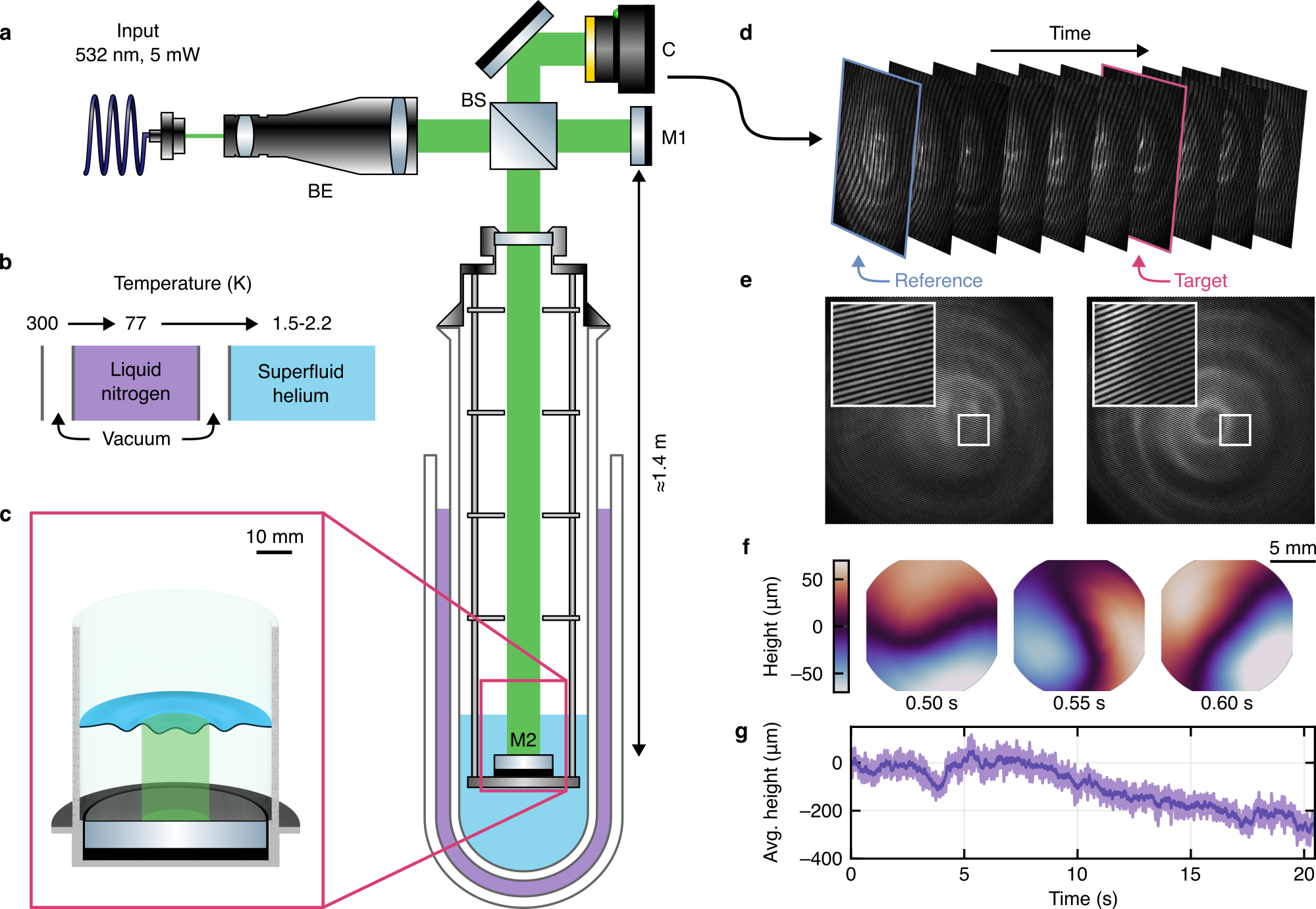}
    \caption{%
    \textbf{Digital holography for helium bath cryostats.}
    \panel{a}~The optical setup employs a laser beam, which is expanded by a beam expander (BE) and split into reference (horizontal) and probe (vertical) beams by a beam splitter (BS). Mirrors M1 and M2 reflect the beams back, forming a Michelson interferometer. The hologram, obtained by recombining the two beams and carrying information about the fluctuations of the superfluid interface, is recorded by a high-speed camera (C). %
    \panel{b}~Thermal profile typical of bath cryostats, including insulating evacuated spaces and a liquid nitrogen jacket. %
    \panel{c}~Half-section view of the experimental zone. The size and approximate location of the expanded laser beam are indicated by green shading. %
    \panel{d}~A series of camera frames shows variations in the acquired holographic pattern. %
    \panel{e}~To reconstruct the superfluid interface topography, each target frame is digitally compared with the reference frame (see Appendix~\ref{sec:dh} for details). %
    \panel{f}~Snapshots of the interface height display an oscillating surface wave. The images are cropped to match the area sufficiently illuminated by the laser beam. %
    \panel{g}~Average interface height as a function of time, revealing the gradual evaporation of superfluid helium. The shaded region indicates one standard deviation ($1\sigma$) interval.}  
    \label{fig:scheme-wet}
\end{figure*}

We recently demonstrated~\cite{S_Barroso2023-zh} that a compact DH setup can resolve surface waves in water with amplitudes as small as $10$~nm over a field of view of a few square centimetres. However, DH imaging has yet to be applied to free-surface measurements in superfluid helium. This is due to a number of technological challenges, particularly concerning the coupling of the probe beam to the superfluid sample. The cryogenic environment necessary to maintain helium in the superfluid (or even liquid) phase typically constrains the admissible optical power, and requires long free-space beam paths, coherent bundles of optical fibers~\cite{skachko2026} or the development new approaches to compensate for the thermal contraction of optical components kept at low temperatures~\cite{eikelman2026}. Crucially, mechanical vibrations inherent to cryogenic systems, arising, e.g. from boiling cryogens, vacuum pumps and cryocoolers, raise concerns about the feasibility of high-precision optical measurements. 

Here, we present the implementation of off-axis DH as a minimally invasive optical readout scheme for the dynamics of surface waves on superfluid helium interfaces. Our results were obtained in two complementary cryogenic systems: traditional helium bath cryostats and the rapidly advancing technology of cryogen-free refrigerators. In the former setup, the object beam enters the superfluid sample, reflects off a submerged mirror, and retraces its incident path, resembling the configuration of a Michelson interferometer (Fig.~\ref{fig:scheme-wet}a). Meanwhile, for the latter, the object beam is transmitted through the sample, as in a Mach-Zehnder interferometer (Fig.~\ref{fig:scheme-dry}a). We benchmark our reconstructions of the superfluid interfaces by recovering the expected dispersion relation of small-amplitude surface waves, using the time-resolved measurements and machine-learning methods to support our results statistically.

\section{\label{sec:setup}Results}

\subsection{\label{sec:glass}Helium bath cryostat}

A helium bath (``wet'') cryostat operates by immersing an experiment in liquid helium. These systems are mechanically simple and offer excellent temperature stability due to large helium content, typically on the order of litres. The temperature, typically starting at $4.2$~K (boiling point at normal atmospheric pressure), can be further reduced through evaporative cooling by lowering the pressure above the bath, resulting in the cooling power on the order of 1~W \cite{Pobell2007}. In our setup shown in Fig.~\ref{fig:scheme-wet}a, the wet cryostat consists of two double-walled glass Dewar flasks. The inner vessel contains liquid helium and is partially immersed in an open bath of liquid nitrogen (nitrogen jacket), shielding the helium bath from room-temperature blackbody radiation. This fully transparent configuration reaches a base temperature of approximately $1.55$~K, well below helium's superfluid transition temperature of $2.17$~K. Because the flasks' glass walls are curved and vary in thickness, we access the superfluid from above via a vacuum-compatible optical flat.

Optical components required for DH are mounted on a breadboard above the cryostat. A fibre-coupled continuous-wave laser beam first passes through a beam expander, increasing its waist to approximately $17$~mm. The expanded beam is split by a 50:50 beam splitter into reference and probe beams. The reference beam reflects off a static mirror, while the probe beam (approximately $2$~mW incident power) enters the cryostat, traverses the superfluid sample, and reflects off another mirror before recombining with the reference beam at the beam splitter. This design results in a Michelson interferometer with a probe-to-reference arms aspect ratio of approximately 30:1 (see Appendix~\ref{app:exp-details} for details).

Approaching the superfluid sample from above aligns well with the typical construction of wet cryostats. The horizontal temperature gradient is considerably steep (see Fig.~\ref{fig:scheme-wet}b), while the temperature varies more gradually in the vertical direction. This ensures that the cryostat's top steel lid and the optical flat remain stably above the dew point, preventing frost or condensation on the window. The experimental region of interest is a cylindrical enclosure filled with superfluid helium. This experimental cell is suspended from the cryostat lid on fibreglass rods with low thermal conductivity and fitted with evenly spaced annular baffles, which enhance its mechanical stability but also reduce the large-scale convection of gaseous helium within the Dewar, thereby reducing unwanted heat load, as well as potential phase fluctuations of the laser field. The inner radii of the baffles and the diameter of the optical flat ultimately limit the field of view from the top, and hence the width of the collimated expanded beam probing the \mbox{sample}. 

The optical scheme can be aligned for off-axis DH by mounting the room-temperature components on three-adjuster kinematic mounts, eliminating the need to finely adjust the mirror inside the cryostat. With adjustments in the reference mirror (M1 in Fig.~\ref{fig:scheme-wet}a), we can introduce a relative tilt between the optical paths of the reference and probe beams, as crucially required by off-axis DH. The resulting hologram, capturing information about the fluctuating superfluid interface, is recorded by a high-speed CMOS camera. As illustrated in Fig.~\ref{fig:scheme-wet}c, the experimental cell has a diameter of $54$~mm, significantly larger than the 17-mm waist of the laser beam. As a result, the spatial extent of the interface accessible to DH is limited, restricting the region over which surface fluctuations can be resolved.

Our interface reconstruction pipeline is illustrated in Fig.~\ref{fig:scheme-wet}d-f and described in Appendix~\ref{sec:dh}. To analyse a time series of camera frames, we digitally compare each target hologram with a reference frame to extract time- and space-resolved phase shifts. In this implementation, the first frame serves as a reference, allowing us to track the evolution of the superfluid interface relative to its state at the start of the recording, while compensating for total (static) phase aberrations and distortions~\cite{colomb2006total}. Exemplary snapshots of the reconstructed interface (Fig.~\ref{fig:scheme-wet}f) reveal the spatial profile of a dominant sloshing wave with an amplitude of approximately $50~\mathrm{\mu m}$, significantly exceeding the detection threshold of DH, also given in Appendix~\ref{sec:dh}. This mode is excited by ambient mechanical noise within the cryostat, primarily arising from two sources: pressure fluctuations induced by the vacuum pump and the boiling of the liquid nitrogen jacket. Together, these sources inject a broadband stochastic noise into the experiment. As we show below, a closer inspection of interface fluctuations in the frequency domain reveals a multitude of noise-driven, superimposed waves, which can be systematically extracted from the data.

Before proceeding to the spectral analysis, we note that DH also enables precise monitoring of helium evaporation. A gradual decrease in interface height results in a global drift in the reconstructed phase. The time evolution of the average interface height relative to the reference is shown in Fig.~\ref{fig:scheme-wet}g. We observe an evaporation rate of approximately $15~\mathrm{\mu m/s}$, in agreement with standard level measurements in wet systems. These conventional approaches, however, require significantly longer timescales and yield only time-averaged estimates of the evaporation rate.

\subsection{\label{ssec:results-wet}Spectral analysis of normal modes}

\begin{figure*}
    \centering
    \includegraphics[scale=1]{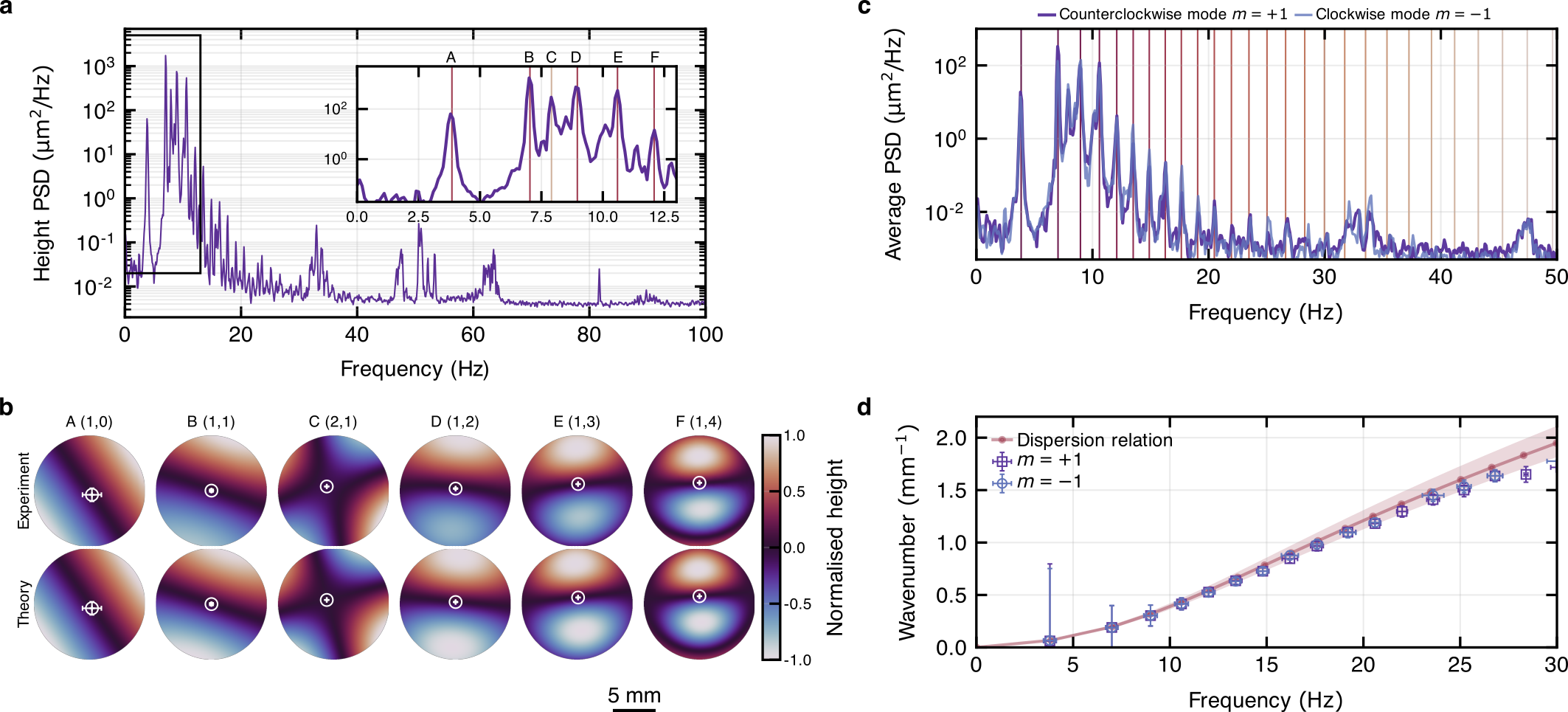}
    \caption{\textbf{Characterisation of superfluid surface waves in a helium bath cryostat.} %
    \panel{a}~Power spectral density (PSD) of interface height fluctuations, showing multiple spectral peaks. Inset: zoom on low-frequency modes A-F, aligned with the predicted Bessel mode frequencies (vertical lines). %
    \panel{b}~Reconstructed spatial profiles of modes A-F (top row) closely match the shape of the corresponding Bessel modes (bottom row). White crosses mark the fitted origin of the polar coordinate system for each mode. %
    \panel{c}~PSD of height fluctuations for one-fold ($m = \pm 1$) modes, overlaid with predicted frequencies of admissible Bessel modes (vertical lines). %
    \panel{d}~Reconstructed dispersion relation (points) agrees well with Eq.~\eqref{eq:disp} (line). Error bars show 1$\sigma$ confidence intervals; the shading around the theoretical relation reflects uncertainty in model parameters. See main text for details.}
    \label{fig:results-wet}
\end{figure*}

A notable property of superfluid helium is its exceptionally low kinematic viscosity $\nu$. At $1.72$~K, the temperature at which our experiments were performed, $\nu = 8.9\times 10^{-9}~\mathrm{m^2/s}$~\cite{donnelly1998}, about 100 times less than water. The characteristic viscous decay time for a surface wave with wavenumber $k = 1~\mathrm{mm^{-1}}$ is $1/(\nu k^2) \approx 113$~s, which significantly exceeds the data acquisition timescale, approximately 20~s. We can therefore treat the fluctuating superfluid interface as a superposition of undamped normal modes, i.e. solutions of the Helmholtz equation in a cylindrically symmetric domain. These modes take the form
\begin{equation}
    \psi_{mn}(r,\phi,t) \propto J_{|m|}(k_{mn}r) \exp(im\phi - i\omega_{mn} t),
    \label{eq:normal}
\end{equation}
where $(r,\phi)$ are polar coordinates, $t$ denotes time, and $m$, $n$ label the azimuthal and radial indices, respectively. The radial dependence of normal modes is given by Bessel functions of the first kind $J_{|m|}$. Due to the finite size of the system, the spectrum of normal modes is discrete, with admissible wavenumbers $k_{mn}$ determined by the boundary condition at $r = R = 27$~mm. Here, we consider a Neumann boundary condition, compatible with an interface freely slipping at the wall and requiring the radial derivative of the mode function to vanish at the boundary, i.e. $J_{|m|}^\prime(k_{mn}R) = 0$. Each normal mode is thus uniquely defined by integers $m$, $n$ ($n \geq 0$) that lead to wavenumber $k_{mn}$, and frequency $\omega_{mn}$. Assuming that the modes are non-interacting and weakly driven, the wavenumbers and frequencies are related by the dispersion relation~\cite{whitham1999},
\begin{equation}
    \omega_{mn}^2 = \left(g + \frac{\sigma}{\rho}k_{mn}^2\right) k_{mn} \tanh\left(h_0 k_{mn}\right),
    \label{eq:disp}
\end{equation}
where $g$ is the gravitational acceleration, $\sigma = 3.2\times 10^{-4}~\mathrm{N/m}$ is the surface tension at $1.72$~K, $\rho = 145~\mathrm{kg/m^3}$ is the density, and $h_0 = 20$~mm is the depth of the superfluid at rest.

We begin our analysis by examining the power spectral density (PSD) of interface height fluctuations, presented in Fig.~\ref{fig:results-wet}a. This spatially averaged frequency spectrum displays a series of well-defined peaks, with the largest amplitudes occurring at low frequencies where viscous damping is minimal. We focus on this region in the panel's inset, where six prominent peaks labelled A-F are identified. First, we note the frequencies of these peaks closely match those of specific normal modes, given by Eq.~\eqref{eq:disp} and displayed as vertical lines. This agreement provides preliminary justification for using normal mode decomposition~\eqref{eq:normal} and Neumann boundary condition in our model. Second, the quality factors of these modes, defined as the ratio of the peak's central frequency to its full width at half maximum, range from 18 to 70. These values indicate damping more significant than expected from viscous dissipation in the superfluid alone, and suggest characteristic decay times on the order of seconds. This additional damping likely arises from the excitation of coupled waves in helium vapour~\cite{bergman1971third}. At $1.72$~K, the kinematic viscosity of the gaseous phase exceeds $\nu$ by a factor of approximately $130$ \footnote{We obtain the kinematic viscosity of gaseous helium, $\nu_g = 1.2\times 10^{-6}~\mathrm{m^2/s}$, as its dynamic viscosity, given by Eq.~(2) in~\cite{nacher1994}, divided by its density, calculated from the standard equation of state of an ideal gas.}, and the excitation of waves in the vapour effectively represents an energy sink. These losses are nonetheless compensated by a stochastic drive from mechanical noise, allowing us to use the full time series of the acquired camera frames to continue with the spectral analysis.

In Fig.~\ref{fig:results-wet}b, we compare the spatial structure of modes A-F with the expected Bessel mode profiles. The observed profiles (top row) are reconstructed by frequency-filtering the height PSD and applying unsupervised machine learning algorithms for signal decomposition, such as Principal Component Analysis~\cite{huamin2017pca} or Truncated Singular Value Decomposition~\cite{manning2008matrix}. These methods extract the most statistically significant (or principal) spatial pattern from a frequency-filtered time series of reconstructed height maps. Conversely, the theoretical profiles (bottom row) are computed as appropriately shifted functional evaluations of $J_{|m|}(k_{mn}r)$ for the $(m,n)$ indices indicated above each plot.

The theoretical predictions accurately replicate experimental data. Each mode exhibits a clear periodic structure centred around an origin point, marked by a white cross. Although the centre of the cylindrical cell, i.e., the physical symmetry centre, is not known a priori, and typically does not coincide with the centre of the field of view, the spatio-temporal resolution provided by DH allows us to recover it from the reconstructed height. This is achieved by fitting the principal spatial profile of each mode to the Bessel mode~\eqref{eq:normal}, and statistically evaluating the most probable origin point (for details and an alternative symmetry determination procedure, see Appendix~\ref{suppl:centre:wet}).

Among the six normal modes identified, only mode C corresponds to $m \neq 1$. Since $m$ defines the wave's azimuthal symmetry, the one-fold modes A, B, and D-F appear as pairs of elevated (red) and depressed (blue) regions. Differences between these modes, encoded in the radial index $n$, become apparent only at larger radii, which lie outside the experimentally accessible region (see Appendix~\ref{suppl:bessel}). Accessing these features would require either further expanding the laser beam or reducing the size of the experimental cell, but both strategies introduce complications. The former increases the complexity of optical alignment, while the latter shifts the normal mode frequencies into a regime where viscous dissipation in the superfluid and gaseous phases become increasingly more important as it scales quadratically with wavenumber. Instead of modifying the experimental layout, we exploit the availability of multiple identified modes. By statistically combining the results of their fitted profiles, we determine the symmetry origin with accuracy better than 2~mm. This process allows us to proceed with the analysis in polar coordinates, focusing on the dominant family of $|m| = 1$ excitations.

The PSDs of the one-fold counterclockwise ($m = 1$) and clockwise ($m = -1$) modes are shown in Fig.~\ref{fig:results-wet}c as overlapping blue lines. To obtain these spectra, we first Fourier transform the interface height along the azimuthal coordinate, where $m$ plays the role of the Fourier conjugate of the angular coordinate $\phi$. We then apply mode-specific filtering and average the resulting signal along the radial axis. The emerging spectral features overlap again with the frequencies of normal modes, indicated by vertical lines for increasing radial index $n$. We are able to resolve up to 16 pairs of modes, with frequencies reaching 30~Hz. This level of detail stands in contrast to Fig.~\ref{fig:results-wet}a, where the identification of higher-frequency peaks is hindered by the rapid growth in the number of possible $(m,n)$ mode combinations~\cite{ajithkumar2025}.

Fitting radial profiles of all mode pairs allows us to identify their wavenumbers and present the dispersion relation of superfluid surface waves in Fig.~\ref{fig:results-wet}d. Experimental data (blue points) show excellent agreement with the expected relation given by Eq.~\eqref{eq:disp} (red line). The shaded region denotes the error band, reflecting uncertainties from the estimated 5-mK temperature fluctuations and a conservative 2-mm margin in determining $R$ and $h_0$. The former arises from unknown thermal contractions of the cell, while the latter occurs due to the limited accuracy of estimating $h_0$ through the side wall of the glass cryostat.

Starting from the readout of a fluctuating superfluid interface height, we have shown how DH enables accurate resolution of a large number of normal modes in both frequency and space, even when their amplitudes are below the micrometre scale (see Fig.~\ref{fig:results-wet}c). We emphasise that the presented methodology is minimally invasive and that knowledge of the experimental cell's shape and approximate size is sufficient to identify numerous modes excited by ambient mechanical noise and to fully characterise their dispersion. Establishing DH in a traditional helium bath cryostat lays the foundation for its broader application in cryogen-free systems, where optical alignment, mechanical noise and heat load due to optical access pose greater challenges. Our implementation of DH under such conditions is addressed in the following section.

\subsection{\label{sec:fridge}Cryogen-free refrigerator}

\begin{figure}
    \centering
    \includegraphics[width=\columnwidth]{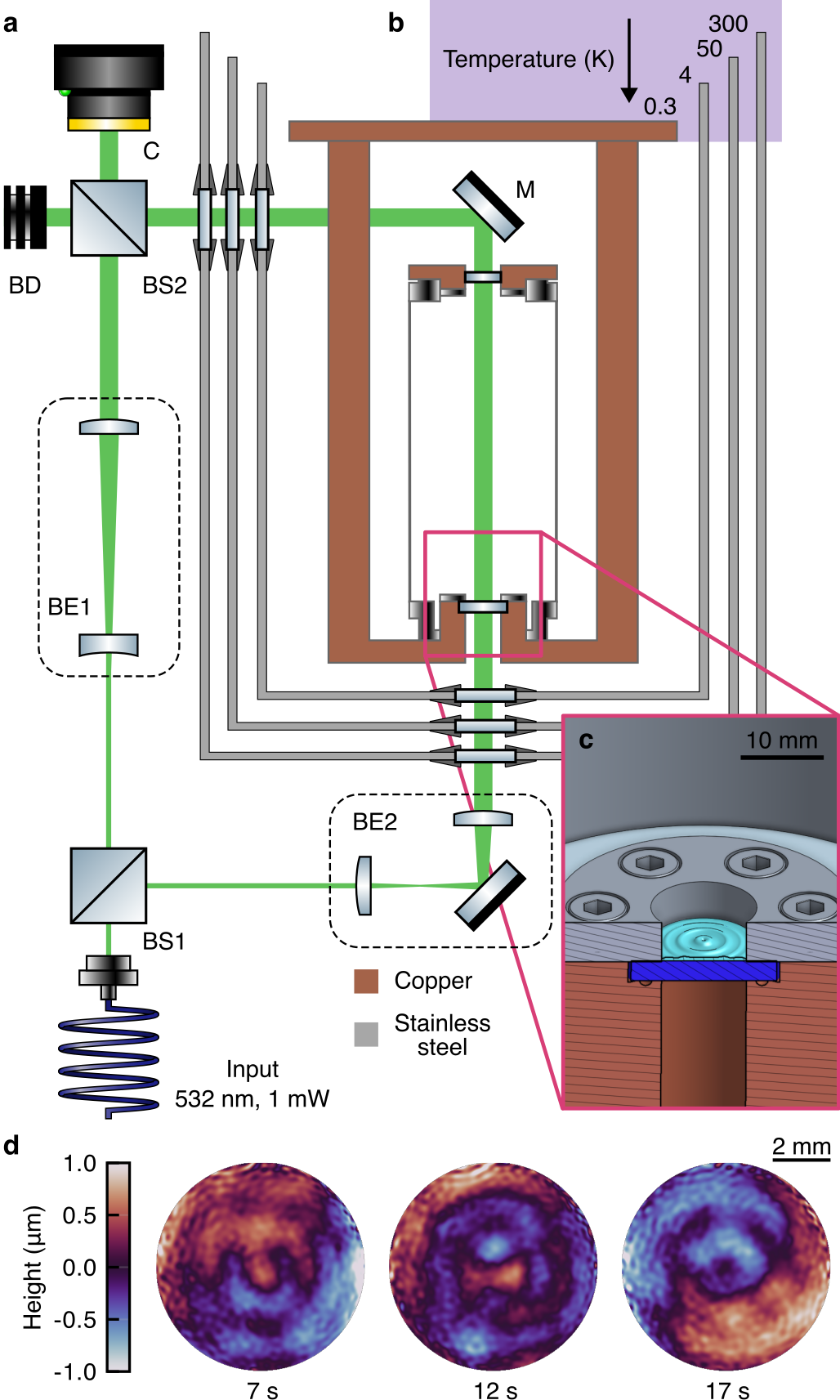}
    \caption{\textbf{Digital holography for cryogen-free refrigerators.} %
    \panel{a}~Mach-Zehnder interferometric setup with the reference (left) and probe (right) beams independently expanded to approximately 7~mm diameter using beam expanders BE1 and BE2. The beams are split by beam splitter BS1 and recombined at BS2. The probe beam travels vertically through the cryostat and is redirected to the horizontal plane by a periscope between mirror M and BS2 (not shown). As in the previous system, the holograms are recorded by a high-speed camera (C). %
    \panel{b}~Temperature map of radiation shields (grey) and the base plate (brown), each coupled to a separate cooling stage. %
    \panel{c}~Section view of the superfluid film coating the bottom optical flat. %
    \panel{d}~Snapshots of nanometre-scale interface excitations. %
    }
    \label{fig:scheme-dry}
\end{figure}

In contrast with their common labelling as ``dry'' refrigerators, this class of low-temperature systems still relies on liquid helium. The latter is, however, confined within closed-cycle loops, eliminating the need for external liquefaction facilities. This design makes them more versatile and increasingly prevalent, particularly in quantum technology and computing applications~\cite{Bluefors2021}. The cooling power in the sample region is typically delivered through a cascade of cooling stages, the first of which is a cryocooler that operates by cyclically compressing and expanding gaseous helium. We use a commercial dry system comprising four cooling stages (Fig.~\ref{fig:scheme-dry}b), reaching a base temperature of approximately 300~mK. The final cooling power of around $100~\mathrm{\mu W}$ is provided by the evaporation of liquid \textsuperscript{3}He by the means of a sorption pump.

\begin{figure*}[ht]
    \centering
    \includegraphics[width=180mm]{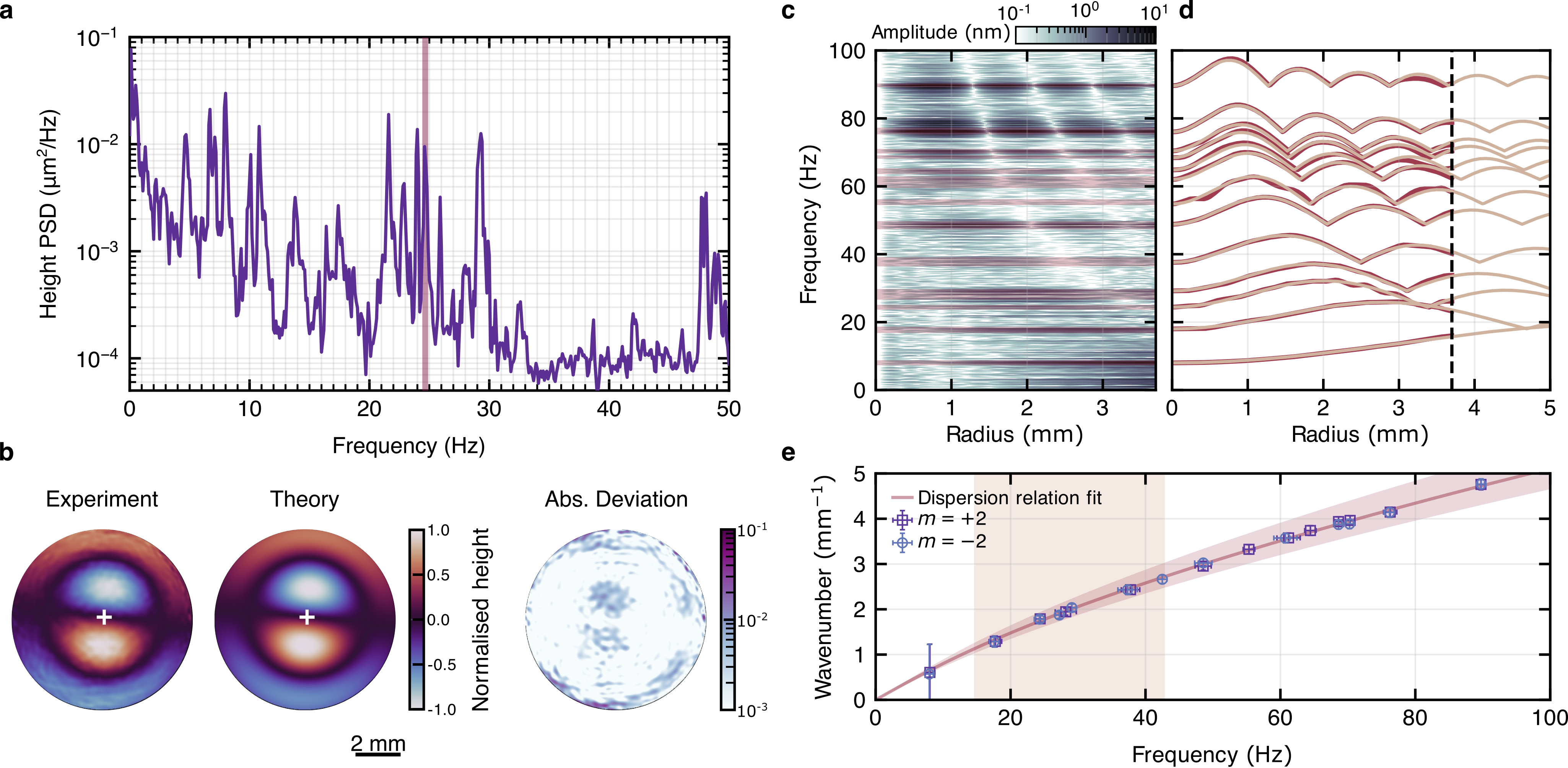}
    \caption{\textbf{Characterisation of superfluid surface waves in a cryogen-free refrigerator.} %
    \panel{a}~Power spectral density (PSD) of interface height fluctuations. The highlighted peak corresponds to the mode selected to determine the symmetry origin. %
    \panel{b}~Reconstructed spatial profile of the mode highlighted in panel \panel{a} (left) with corresponding Bessel mode (centre), and their absolute deviation (right) with total fit residual of $1.6\times 10^{-3}$.  White crosses mark the fitted symmetry origin. %
    \panel{c}~Frequency spectrum of $m = 2$ modes showing standing waves (dark bands) with a distinct nodal structure (white regions). %
    \panel{d}~Reconstructed radial profiles (red) filtered around selected frequencies (red-shaded areas in panel \panel{c}) and fitted with Bessel functions $J_2(k_{2n}r)$ for increasing $n$ (orange). Black dashed line marks the radial extent of our field of view, with the fits extrapolated to the system boundary. The radial profiles are individually normalised, and their starting position along the vertical axis indicates their corresponding frequency. %
    \panel{e}~Identified $m = \pm 2$ modes (points) follow the dispersion relation. Fit to Eq.~\eqref{eq:disp} yields magnification of $1.174 \pm 0.004$ and $h_0 = (578 \pm 21)~\mathrm{\mu m}$ (red line, with the red-shaded area marking the 1$\sigma$ confidence interval). Orange-shaded frequency interval indicates the intermediate regime between shallow-water (lower $f$) and deep-water behaviour (higher $f$). For details, see Sec. \ref{sec:disc}.}
    \label{fig:results-dry}
\end{figure*}

A significant challenge in modern cryogenics is the isolation of mechanical noise generated by their primary cooling mechanism, which introduces substantial vibrations at its operating frequency of 2~Hz and higher harmonics. Mechanical vibrations induce stochastic displacement of the superfluid sample and optical elements located inside the refrigerator. These components shift relative to the room-temperature optics mounted on the optical table that also supports the refrigerator. Although cryogenically-compatible, low-frequency inertial isolation platforms are currently under development to mitigate such effects~\cite{Smetana2024-vh}, our aim is to evaluate the robustness of DH under the original, vibration-prone configuration.

Our proof-of-concept experimental arrangement, schematically illustrated in Fig.~\ref{fig:scheme-dry}a, employs a cylindrical sample cell containing a small quantity of \textsuperscript{4}He. As the system cools to its base temperature, the helium condenses and enters the superfluid phase, forming a relatively thick superfluid film that coats the cell's internal surfaces. Optical access along the cell's axis of symmetry is achieved through a set of sapphire flats, enabling DH in a Mach-Zehnder interferometric configuration. A collimated laser beam is split into reference and probe beams by beam splitter BS1. Each beam is independently expanded (BE1 and BE2), and the beams are later recombined at the beam splitter BS2. As in the previous setup, the resulting hologram is recorded by a high-speed camera  (see Appendix~\ref{app:exp-details} for details). In this configuration, the probe beam (incident power on the order of $100~\mathrm{\mu W}$) may traverse the superfluid film twice, as the film coats each optical port. To avoid this, we heat the top of the cell to $(1.999\pm0.001)$~K, sufficient to remove the top superfluid coating, so that the laser beam interacts with only a single superfluid layer. A detailed view of the interaction zone is shown in Fig.~\ref{fig:scheme-dry}c.

Typical fluctuations of the superfluid interface reconstructed via DH are presented in Fig.~\ref{fig:scheme-dry}d. The amplitude of surface waves is approximately 50 times smaller than in the wet system, while the data appear noisier compared to the corresponding snapshots in Fig.~\ref{fig:scheme-wet}, and feature the characteristic speckle noise of DH~\cite{goodman2007speckle, montresor2016quantitative}. On the other hand, a substantial cross-section of the sample cell (approximately $50\%$) is illuminated by the laser beam.

The spatially-averaged spectrum of these waves, shown in Fig.~\ref{fig:results-dry}a, displays several overlapped spatial modes within each frequency peak, likely due to normal modes with similar oscillating frequencies (see Appendix~\ref{suppl:centre:dry}). This prevents us from employing the procedure presented in Fig.~\ref{fig:results-wet} to multiple low-frequency peaks, as most do not reveal clear spatial profiles despite the quality factors of the 7 highest peaks range from $34$ to $125$. We hence concentrate our analysis on the most prominent one-fold wave, corresponding to mode $(m,n) = (1,2)$, oscillating at $24.6$~Hz with quality factor $109$. By fitting the principal spatial profile using the Bessel mode~\eqref{eq:normal}, we locate the centre of the experimental cell, as presented in Fig.~\ref{fig:results-dry}b. This step enables us to employ the polar coordinate system and separate individual azimuthal modes, i.e., surface waves characterised by a specific periodicity in the azimuthal direction, from the height fluctuation field.

In Fig.~\ref{fig:results-dry}c, we illustrate our approach by plotting the amplitude spectrum of $m = 2$ excitations as a function of radius. The waves are excited only in specific frequency bands (dark regions) and display a distinct radial structure, with nodes appearing as white gaps. By averaging over the red-shaded frequency intervals, we reconstruct the radial amplitude profiles of these modes in Fig.~\ref{fig:results-dry}d. The nodal structure becomes even more pronounced when the experimental data (red lines) are fitted with Bessel functions $J_{2}(k_{2,n}r)$ for increasing $n$ (orange lines). These fits are extrapolated beyond the accessible field of view, limited to a radius of approximately $3.6$~mm, towards the physical boundary of the superfluid sample, equal to 5~mm. Extrapolated curves clearly illustrate how higher-frequency modes exhibit an increasing number of nodes in the radial direction.

To confirm that the observed excitations correspond to surface waves propagating on the superfluid film, we combine spatial and temporal information to reconstruct the underlying dispersion relation, displayed in Fig.~\ref{fig:results-dry}e. We specifically extract the frequencies and wavenumbers of all two-fold modes ($m = \pm 2$, coloured points) and fit the dispersion relation (Eq.~\eqref{eq:disp}, red line) as follows. We fix the surface tension-density ratio to $\sigma/\rho = 2.43\times 10^{-6}~\mathrm{m^3/s^2}$, corresponding to temperature of $(540\pm3)$~mK~\cite{donnelly1998}, and introduce two fitting parameters: the film thickness $h_0$ and a constant prefactor that rescales the wavenumbers extracted from Fig.~\ref{fig:results-dry}d to account for the holograms' magnification. The latter parameter was introduced to adjust the divergence created by the superfluid sample acting as a plano-concave lens due to its meniscus with the wall (see Appendix, Fig.~\ref{fig:s:dry-comparison}). The best fit, yielding magnification of $1.174 \pm 0.004$ and $h_0 = (578 \pm 21)~\mathrm{\mu m}$, accurately describes the observed excitations up to 90~Hz. Note that consistent results are obtained by analysing other azimuthal numbers (see Appendix~\ref{suppl:extended}), further supporting our interpretation of the observed fluctuations as surface waves propagating on a sub-millimetre-thick superfluid film and hence demonstrating that DH is applicable even in mechanically noisy cryogenic systems.

Lastly, we remark that Fig.~\ref{fig:results-dry}e indicates that positive- and negative-$m$ modes are frequency degenerate. This contrasts, e.g., with \cite{ellis1993quantum,Sachkou2019-jk}, where the observed splitting of counter-propagating mode pairs is associated with the presence of persistent superflows, a superfluid analogue to dissipationless currents in superconducting coils. Although this effect may, in principle, also occur in the dry system (see Appendix~\ref{suppl:vortex}), the induced frequency splitting is likely too subtle to be presently resolved.

\section{\label{sec:disc}Discussion and Conclusion}

Minimally invasive readout of fluid interfaces is essential for the experimental investigation of wave dynamics in fluids. Our results demonstrate that off-axis DH offers a robust and versatile optical technique for this purpose. We successfully implemented off-axis DH in the challenging cryogenic environment required to maintain helium in its superfluid phase, enabling full-field reconstruction of surface excitations down to sub-kelvin temperatures. In sufficiently simple geometries, the method provides quantitative access to the decomposition of stochastically driven surface noise into individual normal modes. By extracting the corresponding frequencies and wavenumbers, we reconstructed the dispersion relation of gravity-capillary waves in both wet and dry cryogenic systems.

The dispersion relation is inherently nonlinear due to the finite thickness (or height) of the superfluid layer. However, for wavelengths much longer than $h_0$, i.e., at sufficiently low frequencies, wave propagation occurs in the so-called shallow-water limit, where the dispersion relation becomes linear, $\omega_{mn} = ck_{mn}$, with $c = \sqrt{gh_0}$ denoting the constant wave propagation speed. Such nondispersive and minimally damped shallow-water waves in superfluid helium constitute a valuable platform for realising analogue gravity systems~\cite{schutzhold2002} and for testing predictions of quantum field theory, such as the analogue version of the Unruh effect~\cite{Bunney2024-wv}, thereby bridging concepts from condensed matter and high-energy physics.

In the dataset presented in Fig.~\ref{fig:results-wet}, we find $c = (456 \pm 22)~\mathrm{mm/s}$, but the linear approximation breaks down below the lowest-frequency mode detected (cf. Fig.~\ref{fig:results-wet}d). In superfluid films, however, the propagation speed is significantly reduced. In the example shown in Fig.~\ref{fig:results-dry}, we measure $c = (75 \pm 2)~\mathrm{mm/s}$, which effectively extends the shallow-water regime to higher frequencies. This extent is illustrated in Fig.~\ref{fig:results-dry}e by the orange-shaded region: the shallow-water approximation holds below this band, and we successfully resolved modes that fall within this range. Conversely, at frequencies above the shaded region, the waves are governed by deep-water dynamics, where the dependence on the fluid thickness becomes negligible ($\tanh(h_0k) \approx 1$ for all $k$), and the dispersion relation is determined by the combined effects of gravity and surface tension. The orange region marks, in fact, the crossover between these two limits, where neither approximation is sufficient, and the full dispersion relation~\eqref{eq:disp} must be used. We define the boundaries of this intermediate regime as the frequency range where the shallow- and deep-water approximations deviate by more than $10\%$ from the full expression.

Besides dispersion, another relevant effect influencing wave dynamics is nonlinearity. For the shallow-water waves in cylindrical geometry, this is quantified by the Ursell number \cite{ursell1953}, $\mathrm{Ur} = h_{mn}\lambda_{mn}^2/h_0^3$, where $h_{mn}$ denotes the amplitude of the Bessel mode $(m,n)$, $\lambda_{mn}$ is its wavelength, and $h_0$ is the fluid depth at rest. For the wet system, we obtain the upper estimate of $\mathrm{Ur}$ by considering the longest-wavelength $(1,0)$ mode, with $\lambda_{1,0} = 92$~mm, and $h_{1,0} \approx 50~\mathrm{\mu m}$ estimated from Fig.~\ref{fig:scheme-wet}f. For $h_0 = 20$~mm, we obtain $\mathrm{Ur} = 0.05$, indicating that all low-frequency modes discussed in Sec.~\ref{ssec:results-wet} are well described by linear theory, valid for $\mathrm{Ur} \ll 1$.

For the dry system, using $h_{1,0} = 1~\mathrm{\mu m}$ (Fig.~\ref{fig:scheme-dry}d), $\lambda_{1,0} = 17$~mm, and $h_0 = 578~\mathrm{\mu m}$, the Ursell number is $\mathrm{Ur} = 1.5$, suggesting weak nonlinearity. However, the spectral decomposition in Fig.~\ref{fig:results-dry}c shows that the analysed modes typically oscillate with amplitudes about 100 times smaller than the estimated wave amplitude, placing the considered wave modes to the linear regime.

In order to further extend the frequency range over which nondispersive dynamics apply, the thickness of the superfluid layer must be reduced. Although holographic imaging of such thin superfluid films is left for future work, we note that surface waves can propagate in superfluid films down to nanometre-scale thickness. In this regime, their dynamics is affected by interactions between the superfluid and the substrate~\cite{Sfendla2021-pr}, but since $\mathrm{Ur}\propto h_0^{-3}$, decreasing the film thickness can also allow to access nonlinear regime. DH can therefore complement optomechanical techniques well established to explore these dynamics \cite{reeves2025}.

While our results outline the viability of our approach, it is crucial to identify the main factors currently limiting its performance. The sensitivity of the technique is primarily constrained by phase noise in the hologram acquisition and reconstruction processes~\cite{chen2018phase}. Several sources contribute to such noise, including sensor noise in the camera and fluctuations in the laser phase along the optical path due to mechanical vibrations transmitted to the optical table and cryostat. In our cryogen-free setup, we observed speckle noise more pronounced than in the wet system, which we attribute to the diffraction of the coherent laser beam through multiple optical components, particularly the optical flats (see Fig.~\ref{fig:scheme-dry}a). The randomly varying roughness and losses along the probe optical path are coherently mixed at the camera plane, creating minute random beam interferences, i.e. a speckle wave~\cite{goodman2007speckle}. However, such effects can be reduced via denoising algorithms tailored towards speckle reduction, see e.g.~\cite{bianco2018strategies}, which future implementations should address.

Our optical setups were designed to integrate seamlessly with the available cryogenic infrastructure, without complex imaging lens arrangements, as used in other low-temperature setups, e.g., for imaging trapped atomic clouds~\cite{smits2020imaging}. In contrast to advanced DH schemes developed for metrology~\cite{huang2024quantitative}, often targeting diffraction-limited sample resolution and relying on more elaborate optical configurations, our implementation prioritised operating flexibly over long optical path lengths and large regions of interest. The detection of sub-micrometre-scale surface waves confirms that DH can be effectively deployed in other low-temperature laboratories without the need for major modifications to existing equipment.

In summary, we have demonstrated a cryogenically compatible implementation of off-axis digital holography for broadband, spatio-temporal imaging of free-surface waves in superfluid helium. This minimally invasive and versatile optical readout enables high-resolution characterisation of surface excitations in both bulk and film configurations. By establishing digital holography as a practical and precise technique in cryogenic environments, our results open new pathways for next-generation experiments, ranging from surface-mediated quantum turbulence~\cite{sultanova2023generation} and vortex-wave coupling~\cite{peretti2023direct}, to coherent dynamics in superfluid helium-based quantum field theory simulators~\cite{jarema2025information}.

\section*{Acknowledgements}
VSB, P\v{S}, SS, PS, RP, AJK, and SW extend their appreciation to Science and Technology Facilities Council on Quantum Simulators for Fundamental Physics (ST/T006900/1) as part of the UKRI Quantum Technologies for Fundamental Physics programme. VSB, CG, ID, LS, MC, and SW gratefully acknowledge the support of the Leverhulme Research Leadership Award (RL2019-020). SCA is supported by the Engineering and Physical Sciences Research Council (EP/T517902/1). SW also acknowledges the Royal Society University Research Fellowship (UF120112, RF/ERE/210198, RGF/EA/180286, RGF/EA/181015). Schemes of optical set-ups in this work include components obtained from the \emph{gwoptics Component Library} created by Alexander Franzen.

\section*{Author contributions}
VSB, SCA, AJK, and SW developed, and VSB, SCA, and RP implemented the DH methodology. P\v{S}, CG, ID, SS, PS, LS, and MC integrated the optical readout with wet and dry superfluid systems. VSB analysed the data. VSB and P\v{S} wrote the paper, with contributions from all authors. AJK and SW supervised the project. P\v{S} is the corresponding author.

\section*{Conflict of interest}
The authors declare that they have no competing interests.

\section*{Data availability}
The data that supports the findings of this study are available from the corresponding author upon reasonable request.
\vfill
\clearpage

\appendix
\onecolumngrid

\section{\label{app:exp-details}Experimental and numerical parameters}

\subsection{Helium bath cryostat}

In the optical configuration of Fig.~\ref{fig:scheme-wet}a, the beam expander BE provides 20 times magnification to the input Gaussian beam, yielding a width ($1/e^2$ diameter) of $17~\mathrm{mm}$ with nominal full-angle divergence below $20~\mathrm{\mu rad}$. The beam propagates between the sample and detection planes over $d\approx1.6~\mathrm{m}$, where it illuminates the camera sensor (Phantom VEO~640L). Due to the beam's negligible divergence and deflection from the propagation axis, we argue that numerical propagation of the wavefront (see Eq.~\eqref{eq:beam_propagation} below) is not required to accurately reconstruct the sample height profile. We confirmed this assumption by repeating the analysis of the reconstructed sample height with the numerically propagated wavefront (see Fig.~\ref{fig:s:wet-propagated}). The results display excellent agreement with those presented in Figs.~\ref{fig:scheme-wet} and~\ref{fig:results-wet}, where the wavefront was not propagated back to the sample plane.  

We acquired $20.48~\mathrm{seconds}$-long datasets of $4096$ images with $1536\times1536~\mathrm{pixel^2}$ sampled at $200$ frames per second. The spatially averaged PSDs in Fig.~\ref{fig:results-wet}a,c were computed using Welch's method~\cite{welch1967use}, with $10~\mathrm{s}$-long Hann windows overlapped every $4~\mathrm{s}$. To extract the time series of the individual modes in Fig.~\ref{fig:results-wet}b, we employed fifth-order Butterworth bandpass filters~\cite{harris1978use} centred around each of the peaks in Fig.~\ref{fig:results-wet}a, with the filter width determined by the full width at half maximum of the corresponding peak.

\subsection{Cryogen-free refrigerator}
\begin{figure*}[h!]
    \centering
    \includegraphics[scale=1]{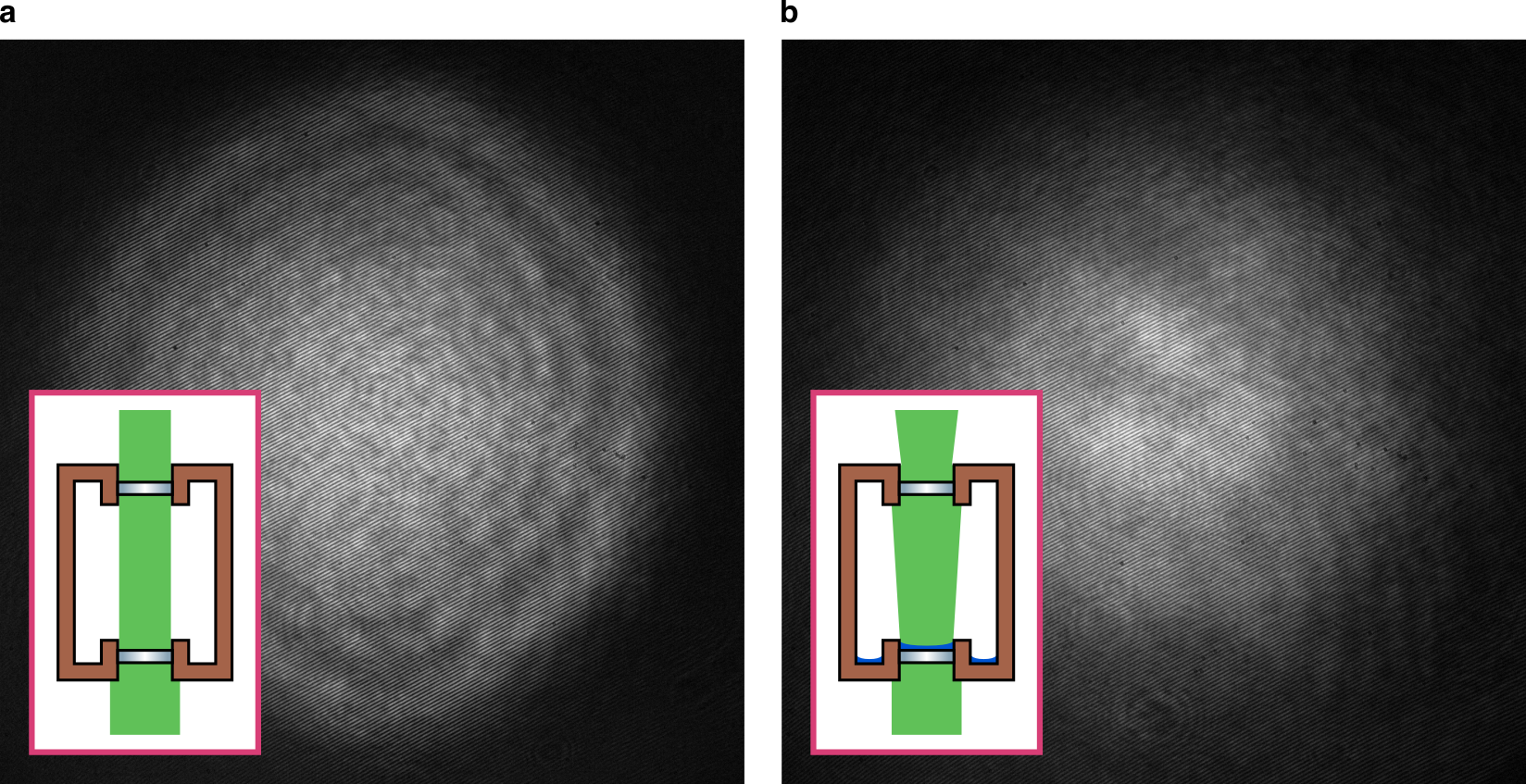}
    \caption{\textbf{Effect of the superfluid film on digital holograms in the cryogen-free refrigerator.} %
    \panel{a}~Image acquired by keeping the experimental cell with helium gas above the superfluid transition temperature. Diffraction rings are visible due to the probe beam grazing through both optical ports.
    \panel{b}~Interference pattern when superfluid helium film is condensed and pooled on the bottom optical port. Evidently, the presence of the film introduces a divergence in the probe beam, which is likely due to the wetting of the walls creating a curved superfluid interface akin to a plano-concave lens as schematically indicated in the figure inset.}
    \label{fig:s:dry-comparison}
\end{figure*}

In the optical configuration of Fig.~\ref{fig:scheme-dry}a, the beam expander BE2 provides 16 times magnification to the input probe Gaussian beam, yielding a width ($1/e^2$ diameter) of $14~\mathrm{mm}$ with nominal full-angle divergence below $30~\mathrm{\mu rad}$. The beam propagates between the sample and detection planes over $d\approx1.7~\mathrm{m}$, where it illuminates the camera sensor (Phantom Miro Lab 340). The beam is cropped by the optical port in Fig.~\ref{fig:scheme-dry}c with $10~\mathrm{mm}$ diameter, introducing diffraction patterns and increasing the divergence of the object wave. A comparison between holographic images of the system with and without superfluid helium (see Fig.~\ref{fig:s:dry-comparison}) reveals that the beam becomes further divergent in the presence of a film. Accordingly, we employ numerical wavefront propagation (see below), as commonly required by off-axis DH, to correct for such complications along the probe beam optical path. 

We acquired $20~\mathrm{seconds}$-long datasets of $4000$ images with $1024\times1024~\mathrm{pixel^2}$ sampled at $200$ frames per second. The spatially averaged PSDs and the frequency filtering required for the normal mode analysis were computed using the same algorithms and parameters as for the helium bath cryostat.

\section{\label{sec:dh}Surface profile reconstruction}

In both interferometric setups discussed here, namely reflective (Michelson) and transmissive (Mach-Zehnder), a digital hologram is formed by coherently mixing two monochromatic laser beams. The object (or probe) wave $E_{\text{obj}}(\bm{x})\exp(i\bm{k}_{\text{obj}}\cdot\bm{x})$ is diffracted by the sample into the sensor (camera) plane, where it interferes with a reference wave $E_\text{ref}(\bm{x})\exp(i\bm{k}_{\text{ref}}\cdot\bm{x})$, in which $\bm{k}_{\text{obj}}$, $\bm{k}_{\text{ref}}$ are the wavevectors of the incoming fields and $\bm{x}=(x,y)$ are the coordinates of the sensor's $x$-$y$ plane at $z=0$. In off-axis configuration, the reference and object waves propagate nearly parallel to the $z$-direction (paraxial propagation) but along slightly misaligned axes (off-axis), creating a wavenumber difference, denoted $\bm{\tilde{k}}=\bm{k}_{\text{obj}}-\bm{k}_{\text{ref}}$. It is worth noting that this difference is related to the angular tilts $(\theta_x,\theta_y)$ between the two propagation axes by $\bm{\tilde{k}}=k_0(\sin\theta_x,\sin\theta_y)$, with $k_0=2\pi/\lambda$ being the laser wavenumber. The resulting intensity pattern recorded by the camera may be written as
\begin{align}
    I &\propto \left|E_\text{ref}\exp\left(i\bm{k}_\text{ref}\cdot\bm{x}\right)+ E_{\text{obj}}\exp\left(i\bm{k}_\text{obj}\cdot\bm{x}\right)\right|^2= |E_\text{ref}|^2 + |E_\text{obj}|^2 + E_\text{ref}^*E_\text{obj}\exp(i\bm{\tilde{k}}\cdot\bm{x}) + \mathrm{c.c.},
    \label{eq:hologram}
\end{align}
where c.c. denotes complex conjugate. This pattern is representative of the images recorded in the experiments (see again Fig.~\ref{fig:scheme-wet}d).

\begin{subequations}\label{eq:phase_to_h}
We assume that, at time $t$, the object wave attains a phase shift $\varphi(\bm{x'},t)$ at the sample plane due to a change in the superfluid surface height $h(\bm{x'}, t)$, where $\bm{x'}$ are the coordinates of the sample plane. In the transmissive (Mach-Zehnder) configuration, this phase shift at the sample plane is given by~\cite{bhaduri2014diffraction}
\begin{equation}
    \frac{\varphi_{\text{T}}(\bm{x'},t)}{2\pi}=\Delta n \frac{h(\bm{x'},t)}{\lambda},
    \label{eq:phase_mz}
\end{equation}
where $\Delta n = n-n_0$ is the refractive index contrast between the superfluid $n$ and free space $n_0$. In the reflective (Michelson) case, the beam travels through the sample twice, hence the phase shift at the sample plane reads~\cite{S_Barroso2023-zh}
\begin{equation}
    \frac{\varphi_{\text{R}}(\bm{x'},t)}{2\pi}=2\Delta n \frac{h(\bm{x'},t)}{\lambda}.
    \label{eq:phase_m}
\end{equation}\end{subequations}
In both configurations, the object wave propagates a distance $d$ from the sample to the camera plane and can be written as
    $E_{\text{obj}}(\bm{x},t) = O(\bm{x},t)\exp\left[i\varphi(\bm{x},t)+iW_O(\bm{x},t)\right]$,
where $O$ denotes the absolute value (amplitude) of the object wave and $W_O$ accounts for all phase shifts unrelated to the sample, e.g., local changes in refractive index or other optical components moving along the optical path. Similarly, the time-independent reference wave can be written as $E_{\text{ref}}(\bm{x}) = R(\bm{x})\exp\left[iW_R(\bm{x})\right]$.

The first two terms in Eq.~\eqref{eq:hologram} correspond to the zero-order diffraction contribution, while the third (fourth) is the $+1~(-1)$ order diffraction and includes the complex object wavefront, whose phase contains information about the sample. The Fourier transform of the intensity image~\eqref{eq:hologram} reveals distinct peaks associated with each diffraction order. The spectrum of the zero-order terms is concentrated around the zero frequency. On the other hand, the spectra of the $\pm 1$ orders are displaced from the origin by $\bm{\tilde{k}}$.
Practically, we isolate the $+1$ order term in Eq.~\eqref{eq:hologram} by applying a circular window~\cite{harris1978use} (2D cosine-tapered, or Tukey, with $\alpha=0.1$) centred around $\bm{\tilde{k}}$ with a large enough width to include all relevant spectral features, but small enough to prevent crossing with other diffraction orders~\cite{cuche2000spatial}. This procedure yields a filtered hologram $I_F$, which reads $I_F = RO\exp[i(\varphi+ W_O-W_R+\bm{\tilde{k}}\cdot\bm{x})]$.
The term $W_O-W_R$ describes total phase aberrations and other optical path length contributions that are propagated from the object plane.  
 
In our phase retrieval procedure, a digital reference is generated as an exact plane wave, i.e., $I_{\text{RC}}=\exp(-i\bm{\tilde{k}}\cdot\bm{x})$. Accordingly, the complex wavefront $\Psi_C(\bm{x},t)$ in the camera plane $z=0$ at time $t$ can be reconstructed using
\begin{align}
    \Psi_C(\bm{x},t) &= I_{\text{RC}}(\bm{x})I_{F}(\bm{x},t) = R(\bm{x}) O(\bm{x},t)\exp\left[i\varphi(\bm{x},t)-iW(\bm{x},t)\right],\label{eq:camera_wavefront}
\end{align}
where $W$ accounts for all additional phase shifts unrelated to the sample. In a stack of recorded images, we eliminate phase aberrations from a target phase frame by computing its difference with respect to the first frame, which acts as our reference phase~\cite{colomb2006total}.

The complete phase reconstruction procedure involves propagating the wavefront reconstructed at the camera plane $\Psi_C$ back to the sample plane, denoted $\Psi_S(\bm{x},z)$, at $z=d$. Multiple numerical beam propagation methods are available~\cite{verrier2011off-axis}, as well as optical configurations that require no propagation, such as image-plane DH~\cite{karray2012comparison}. Without loss of generality, we choose the convolution formulation~\cite{li2009digital} and note it can be directly applied to the complex wavefront~\eqref{eq:camera_wavefront} to reconstruct it at the sample plane, as follows,
\begin{equation}
    \Psi_S(\bm{x},d,t) = \mathcal{F}_{\bm{x}}^{-1}\Big\{H\left(\bm{k},d\right)~\mathcal{F}_{\bm{x}}\left[\Psi_C(\bm{x},t)\right](\bm{k})\Big\}(\bm{x}),\label{eq:beam_propagation}
\end{equation}
where $H$ is the angular spectrum transfer function~\cite{li2007diffraction},
\begin{equation}
    H(\bm{k},z) = \exp\left(-ik_0 z \sqrt{1-\frac{|\bm{k}|^2}{k_0^2}}
    \right).
\end{equation}
We stress that, in the convolution formulation~\eqref{eq:beam_propagation}, the propagated object wavefront at $z=d$ is evaluated at the coordinate system $\bm{x}$ of the camera. 

The phase of $\Psi_S$ can be numerically retrieved and related to the superfluid sample height through Eqs.~\eqref{eq:phase_to_h}. We model the reconstructed height as $h_{\text{rec}}(\bm{x},t) = h(\bm{x},t) + h_\text{noise}(\bm{x},t)$, where $h_\text{noise}$ denotes noise and/or phase aberration contributions from $W$ in Eq.~\eqref{eq:camera_wavefront}. The physical height $h$ has distinctive spatial features given by the normal modes~\eqref{eq:normal} at specific frequencies determined by the dispersion relation~\eqref{eq:disp}. On the other hand, the noise term $h_\text{noise}$ may have contributions at isolated frequencies, e.g. at characteristic mechanical resonances, but with trivial spatial profiles. The results presented in Figs.~\ref{fig:results-wet} and~\ref{fig:results-dry} illustrate this principle and demonstrate the reconstruction of individual normal modes.

\paragraph*{\textbf{Sensitivity of DH}}

The method's sensitivity, defined as the smallest height difference that can be reconstructed from the holograms, depends on the precision with which phase fluctuations $\varphi_\mathrm{T}$ and $\varphi_\mathrm{R}$ can be estimated from the greyscale camera images. Under ideal conditions, the difference between absolutely dark and bright pixels corresponds to a $\pi$-phase difference. In practice, however, the contrast of the acquired holograms (see Fig.~\ref{fig:s:dry-comparison}) is limited by uneven laser field intensity and sensor pixel noise. Given these constraints, we estimate the achievable contrast to be $1/8$ of the full 16-bit camera range, resulting in the smallest detectable phase difference $\varphi_{\min{}}/\pi = 1/2^{13}$. Using Eqs.~\eqref{eq:phase_mz} and \eqref{eq:phase_m}, the minimum height difference resolved by DH becomes $h_\mathrm{min} = 1.16$~nm for the Mach–Zehnder configuration and $h_\mathrm{min} = 0.58$~nm for the Michelson configuration.

\section{\label{suppl:centre}Statistical evaluation of the symmetry origin}

\subsection{\label{suppl:centre:wet}Helium bath cryostat -- signal decomposition vs. time series analysis}

The normal mode analysis and determination of the symmetry origin presented in the manuscript rely on machine learning algorithms for decomposing signals into components, e.g. PCA and Truncated SVD, both available in~\cite{scikit-learn}. Here, we describe the procedure employed and offer an alternative statistical analysis that uses the entire time series of reconstructed normal modes instead. 

\paragraph*{\textbf{Principal Component Analysis}}

First, we apply a filter around a selected PSD peak (as shown in the inset of Fig.~\ref{fig:results-wet}a) and retrieve the time series of spatial profiles at that frequency. The filtered reconstructed height $h_{\text{rec},\omega}(t,r,\theta)$ around a frequency $\omega$ should be well approximated by the corresponding normal mode $\psi_{mn}$ in~\eqref{eq:normal} with frequency given by the dispersion relation~\eqref{eq:disp}, i.e., $\omega_{mn}\equiv\omega$. Thus, 
$$h_{\text{rec},\omega}(t,r,\theta)\approx \Re\{h_{mn}(t)\psi_{mn}(t,r,\theta)\}+\eta(t,r,\theta),$$ 
with some amplitude $h_{mn}$ and residual noise $\eta$. Each spatial point of $h_{\text{rec},\omega}$ is then treated as a sample and the temporal values $t$ as features. In this representation, all spatial points share nearly the same time dependence at frequency $\omega$, differing only by a complex multiplicative factor $h_{mn}$. Consequently, the temporal signals across samples are highly correlated. 

\begin{figure*}
    \centering
    \includegraphics[width=\linewidth]{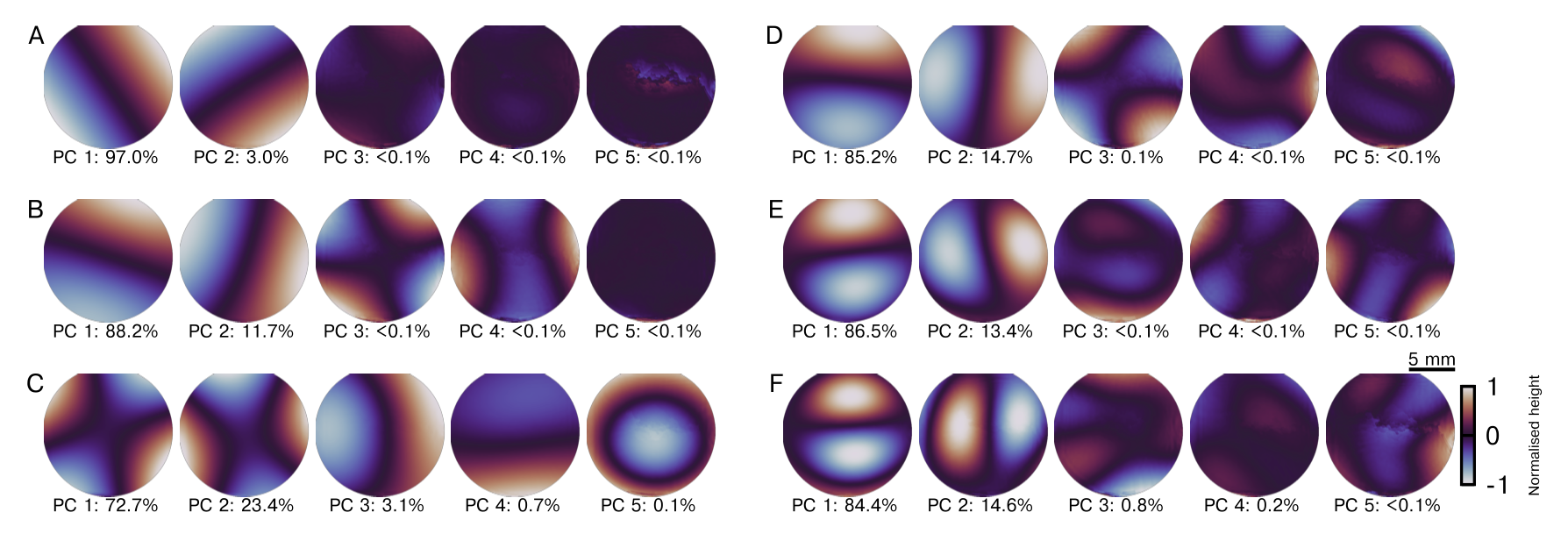}
    \caption{\textbf{Signal decomposition of surface normal modes using PCA.} %
    Each panel (A-F) displays the signal decomposition through PCA of the corresponding normal modes A-F indicated in Fig.~\ref{fig:results-wet}a. The panels show the first five principal components (labelled PC) of the time series of the corresponding peak in the PSD. The labels show the percentage of signal variance explained by each component. Most time series signals can be well-decomposed into the first components with confidences ranging from $72.1\%$ to $97\%$.}
    \label{fig:s:wet-pca}
\end{figure*}

Principal Component Analysis (PCA)~\cite{huamin2017pca} identifies orthogonal directions in feature space (here, temporal patterns) that capture the most significant variance in the dataset. Because the frequency filtering leaves almost all the variance concentrated in a single coherent oscillatory pattern, the first principal component aligns with the dominant temporal waveform at $\omega$. By projecting the signal onto the first principal component, we reconstruct the spatial profile of that mode. Fig.~\ref{fig:s:wet-pca} shows the first five principal components obtained through PCA for the peaks selected in Fig.~\ref{fig:results-wet}a, sorted in descending order of their explained variances, which quantifies the percentage of the signal variance explained by each component. PCA decomposes the time series of all peaks into a dominant principal component explaining more than $85\%$ of the signals in most cases. Note that the second principal components in Fig.~\ref{fig:s:wet-pca} appear to correspond to the modes explained by the first principal components but rotated by 90 degrees.

The first principal component of each PSD peak and its frequency are used to find the corresponding normal mode, as shown in Fig.~\ref{fig:results-wet}b. For that, we perform a least-squares regression of the experimental waveforms with the Bessel functions~\eqref{eq:normal} to estimate the centre of the experimental cell, i.e., our symmetry origin. As part of the fitting procedure, an arbitrary phase is also determined to account for the mode's angular orientation. As shown in Fig.~\ref{fig:s:bessel}, mode A, corresponding to the $(m,n) = (1,0)$ mode, appears nearly as a tilted plane within the field of view (dashed white circle). As a result, any point along the nodal line (dark region) may yield a plausible, but potentially incorrect, estimate of the symmetry centre. Accordingly, the fitting of mode A in Fig.~\ref{fig:results-wet}b presents a strong bias in the determination of the centre of origin.

\begin{figure}
    \centering
    \includegraphics[scale=1]{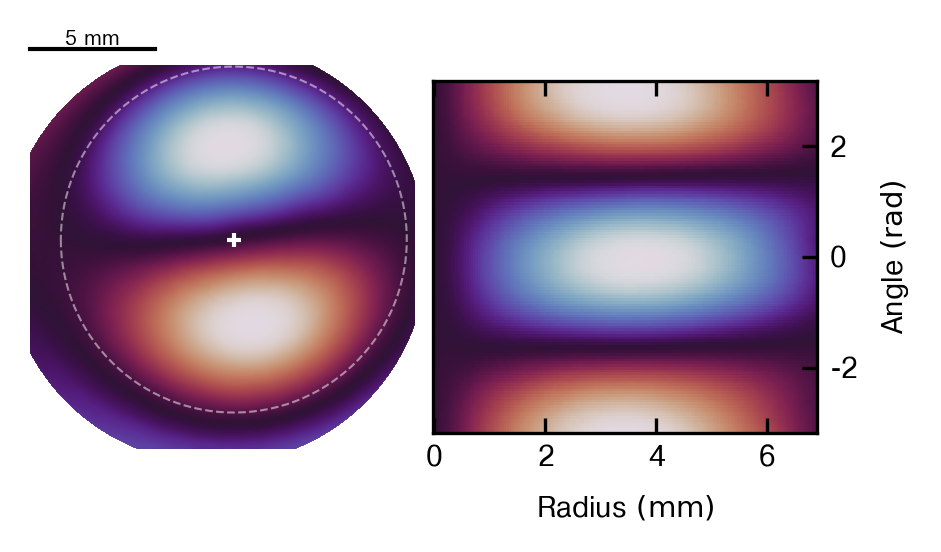}
    \caption{\textbf{Polar transformation of an experimentally reconstructed surface mode.} %
    After identifying the origin of the coordinate system (white cross, left panel), the data within the white dashed circle can be transformed from Cartesian to polar coordinates (right panel).}
    \label{fig:s:polar}
\end{figure}

Accurate determination of the cell's symmetry centre is essential for the subsequent transformation to polar coordinates. Performing this transformation about a misidentified origin leads to apparent mode mixing -- for example, $m = 1$ signals may leak into higher azimuthal orders. As a consistency check, we compare the same waveform plotted in Cartesian (left panel) and polar coordinates (right panel) in Fig.~\ref{fig:s:polar}. The polar plot reveals a clear periodic modulation in the angular direction, validating the transformation and enabling further analyses of the modes' radial structure (see, e.g. Fig.~\ref{fig:results-wet}c and~\ref{fig:results-dry}c).

\paragraph*{\textbf{Time Series Analysis}}

\begin{figure*}
    \centering
    \includegraphics[scale=1]{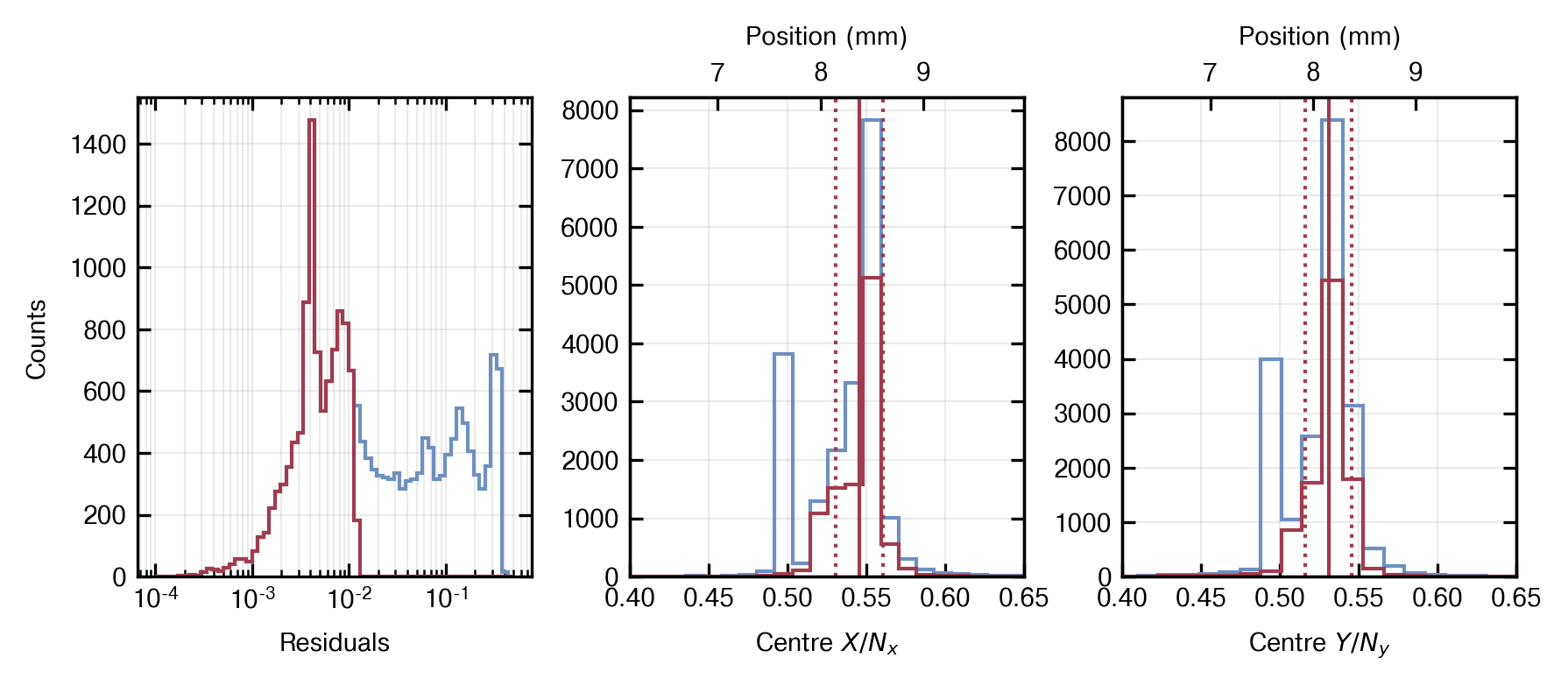}
    \caption{\textbf{Statistical analysis of the fitted centre coordinates.} %
    (Left) Histogram of residuals from two-dimensional Bessel mode fits to the waveforms shown in Fig.~\ref{fig:results-wet}b, obtained by independently fitting all $4095$ target frames for each mode.
    (Middle and Right) Histograms of the fitted origin's horizontal and vertical coordinates $X$ and $Y$, normalised by the image size in pixels ($N_x$ and $N_y$, respectively). Vertical solid lines denote the mean values; dotted lines indicate $1\sigma$ confidence intervals. In all panels, the blue histograms (in the background) show the distributions of all fit results, whereas the red histograms display the second quartile ($50\%$) of residuals, corresponding to fit residuals below the median value of $0.015$.}
    \label{fig:s:distributions}
\end{figure*}

\begin{figure*}
    \centering
    \includegraphics[scale=1]{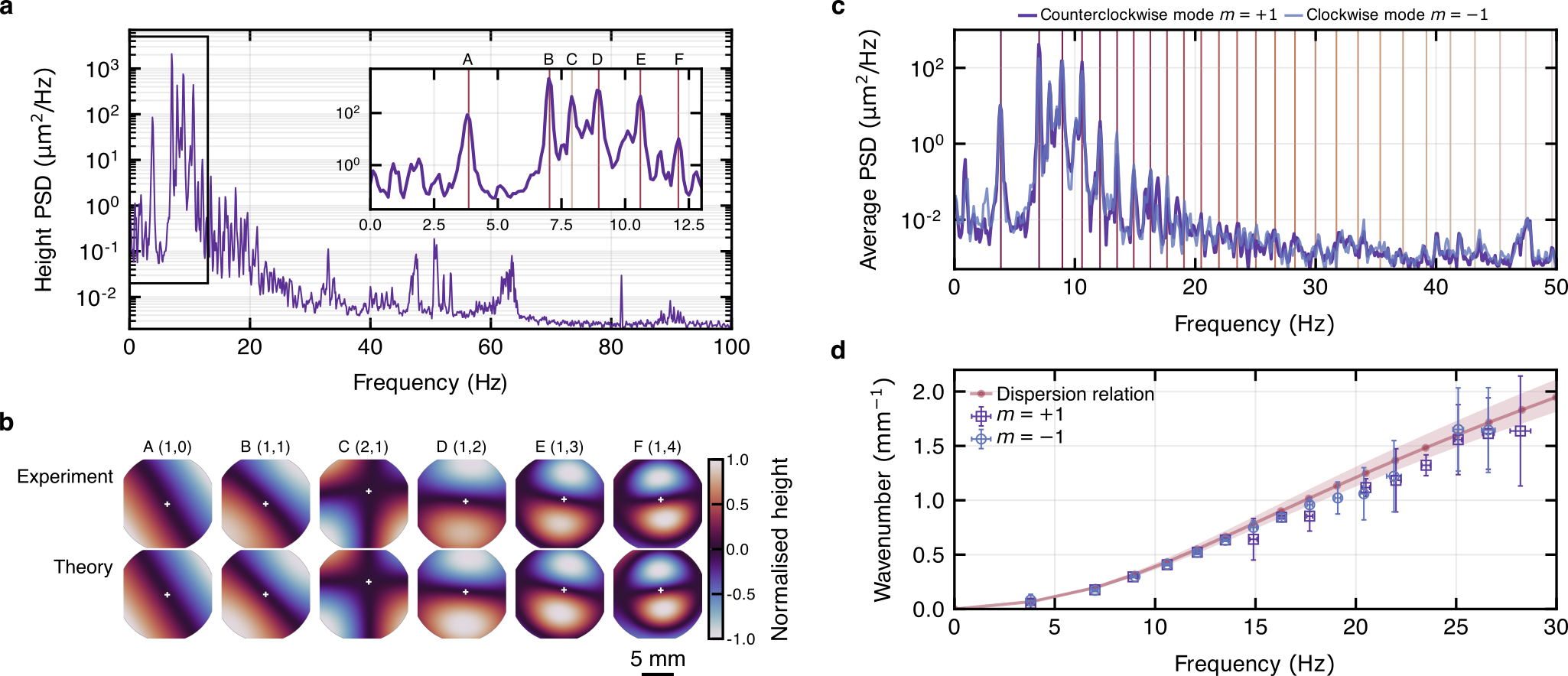}
    \caption{\textbf{Time series analysis of superfluid surface waves in a helium bath cryostat.} %
    \panel{a} Power spectral density (PSD) of interface height fluctuations, showing multiple spectral peaks. Inset: zoom on low-frequency modes A-F, aligned with the predicted Bessel mode frequencies (vertical lines). %
    \panel{b} Reconstructed spatial profiles of modes A-F (top row) closely match the shape of the corresponding Bessel modes (bottom row). White crosses mark the fitted origin of the polar coordinate system for each mode. %
    \panel{c} PSD of height fluctuations for one-fold modes, overlaid with predicted frequencies of admissible Bessel modes (vertical lines). %
    \panel{d} Reconstructed dispersion relation (points) agrees well with Eq.~\eqref{eq:disp} (line). Error bars show 1$\sigma$ confidence intervals; the shading around the theoretical relation reflects uncertainty in model parameters.}
    \label{fig:s:wet-propagated}
\end{figure*}

As an alternative to PCA, we can directly analyse the time series of spatial profiles obtained by filtering around a given PSD peak. To this end, we treat individual camera frames as uncorrelated data samples and employ the same least-squares regression method outlined above in the PCA procedure. In Fig.~\ref{fig:s:distributions}, we statistically evaluate the fitted parameters for the ensemble of six modes over $4095$ time steps. The left panel displays the distribution of fit residuals in arbitrary units (red line). The middle and right panels show distributions of the fitted centre coordinates, together with mean values (solid lines) and confidence intervals (dotted lines). As expected, not all fits converge well at all time steps; thus, we employ a filtering strategy by selecting the best fits with residuals below a certain threshold. Our choice of threshold was the median value (second quartile) of residuals. This post-selection yields narrower histograms for the fitted centre positions, as seen by the red histograms in the middle and right panels of Fig.~\ref{fig:s:distributions}, and allows the determination of the symmetry origin with comparable precision and accuracy as the previous method. For comparison, we show in Fig.~\ref{fig:s:wet-propagated} the same results presented in Fig.~\ref{fig:results-wet}, but using the entire time series instead of principal components.

\subsection{\label{suppl:centre:dry}Cryogen-free refrigerator -- signal decomposition}

\begin{figure*}
    \centering
    \includegraphics[width=\linewidth]{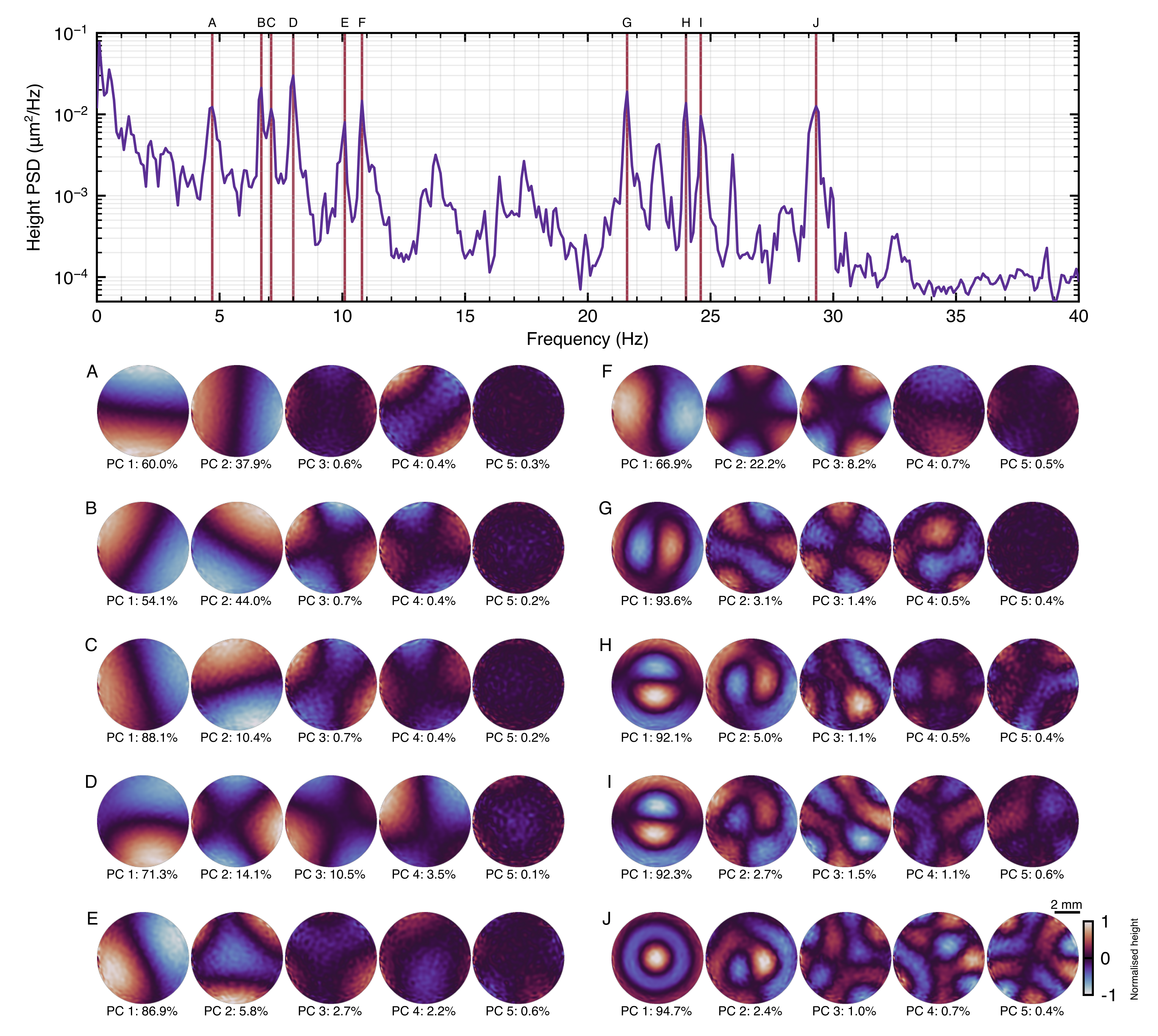}
    \caption{\textbf{Principal component extraction from power spectral density in a cryogen-free refrigerator.} (Top) PSD of interface height fluctuations, analogous to Fig.~\ref{fig:results-wet}a, highlighting the selected normal modes (red lines). (Bottom)
    Each panel (A-J) displays the first five principal components obtained through PCA of the corresponding normal modes A-J indicated in the top panel. The labels show the percentage of signal variance explained by each component. }
    \label{fig:s:dry-pca}
\end{figure*}

In the power spectral density (PSD) of height fluctuations in a cryogen-free refrigerator, shown in the top panel of Fig.~\ref{fig:s:dry-pca}, multiple frequency peaks are visible, similarly to Fig.~\ref{fig:s:wet-pca} obtained from a helium bath cryostat. However, these peaks appear largely overlapped in Fig.~\ref{fig:s:dry-pca} and their amplitudes are much smaller than in Fig.~\ref{fig:s:wet-pca}. This makes the identification of distinct spatial profiles, as in Fig.~\ref{fig:results-wet}b, a challenging task. To circumvent this issue, we employed the PCA procedure outlined in Sec.~\ref{suppl:centre:wet} to the ten most prominent peaks in the PSD (indicated in the top panel of Fig.~\ref{fig:s:dry-pca}). Their first five principal components are shown in the lower panels of Fig.~\ref{fig:s:dry-pca}. It is evident that the first components resemble the expected Bessel-like profiles; however, only modes I and J accurately fit to the normal modes~\eqref{eq:normal}. This is partly due to an extra fitting parameter, the wavenumber $k$ of the normal mode, which increases the complexity of the least-squares regression. Consequently, the fitting procedure only yields trustworthy results for reconstructed spatial profiles that the model describes satisfactorily. For the results presented in Fig.~\ref{fig:results-dry}, we conducted the same analysis of Sec.~\ref{suppl:centre:wet} using mode I only.
\clearpage

\section{\label{suppl:bessel}Spatial profile of Bessel modes}

\begin{figure*}
    \centering
    \includegraphics[scale=0.9]{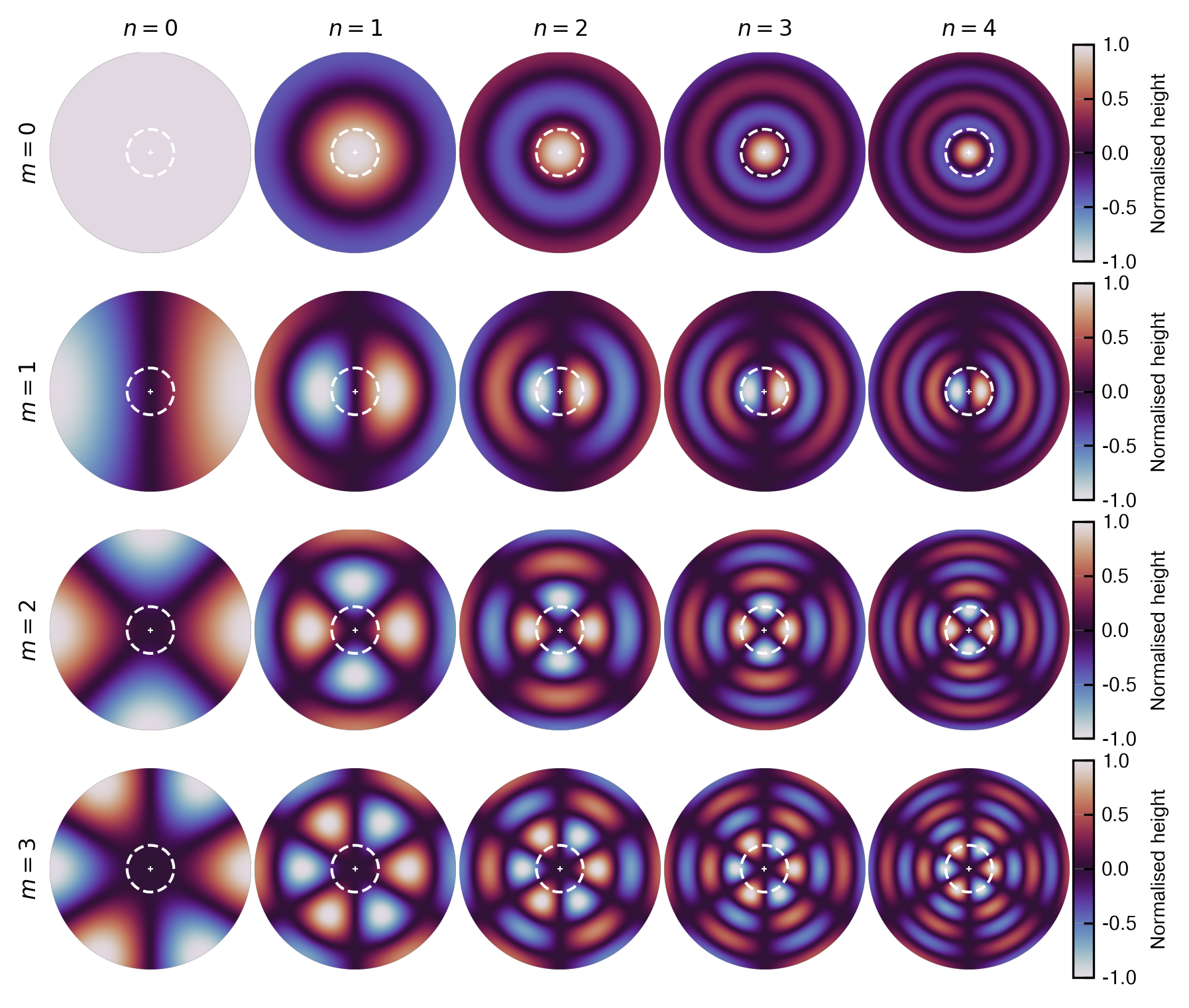}
    \caption{\textbf{Predicted spatial profiles of cylindrical modes with full-slip boundary condition.} %
    Spatial profiles for different normal modes expected in the sample cell submerged in the helium bath cryostat. Dashed white circles denote the accessible field of view for the polar transformation with $6.3$~mm radius.}
    \label{fig:s:bessel}
\end{figure*}

The distinction between different Bessel modes is most evident when considering the full spatial domain in which they are excited. In Fig.~\ref{fig:s:bessel}, we display the expected mode patterns on the superfluid interface, defined in polar coordinates as $J_m(k_{mn}r)\cos(m\phi)$. The wavenumbers $k_{mn}$ are calculated using a Neumann boundary condition at $R = 27$~mm. To illustrate the limited spatial extent accessible in our experiments, a white circle with a $6.3$~mm radius is overlaid at the centre of each panel. Within these circles, the data can be transformed into polar coordinates, as in Fig.~\ref{fig:s:polar}. The information contained within this restricted field of view is consistent with the reconstructed waveforms presented in Fig.~\ref{fig:results-wet}b of the main text. 
\clearpage

\section{\label{suppl:extended}Extended reconstruction of the dispersion relation}

In Fig.~\ref{fig:results-dry}e, we showcase how the dispersion relation of waves propagating on the surface of a superfluid film can be reconstructed from space and time-resolved holographic data. The polar coordinate transformation, outlined in Sec.~\ref{suppl:centre}, allows us to isolate modes with different periodicity along the angular coordinate, i.e., distinct $m$-modes.

\begin{figure*}[h!]
    \centering
    \includegraphics[scale=0.85]{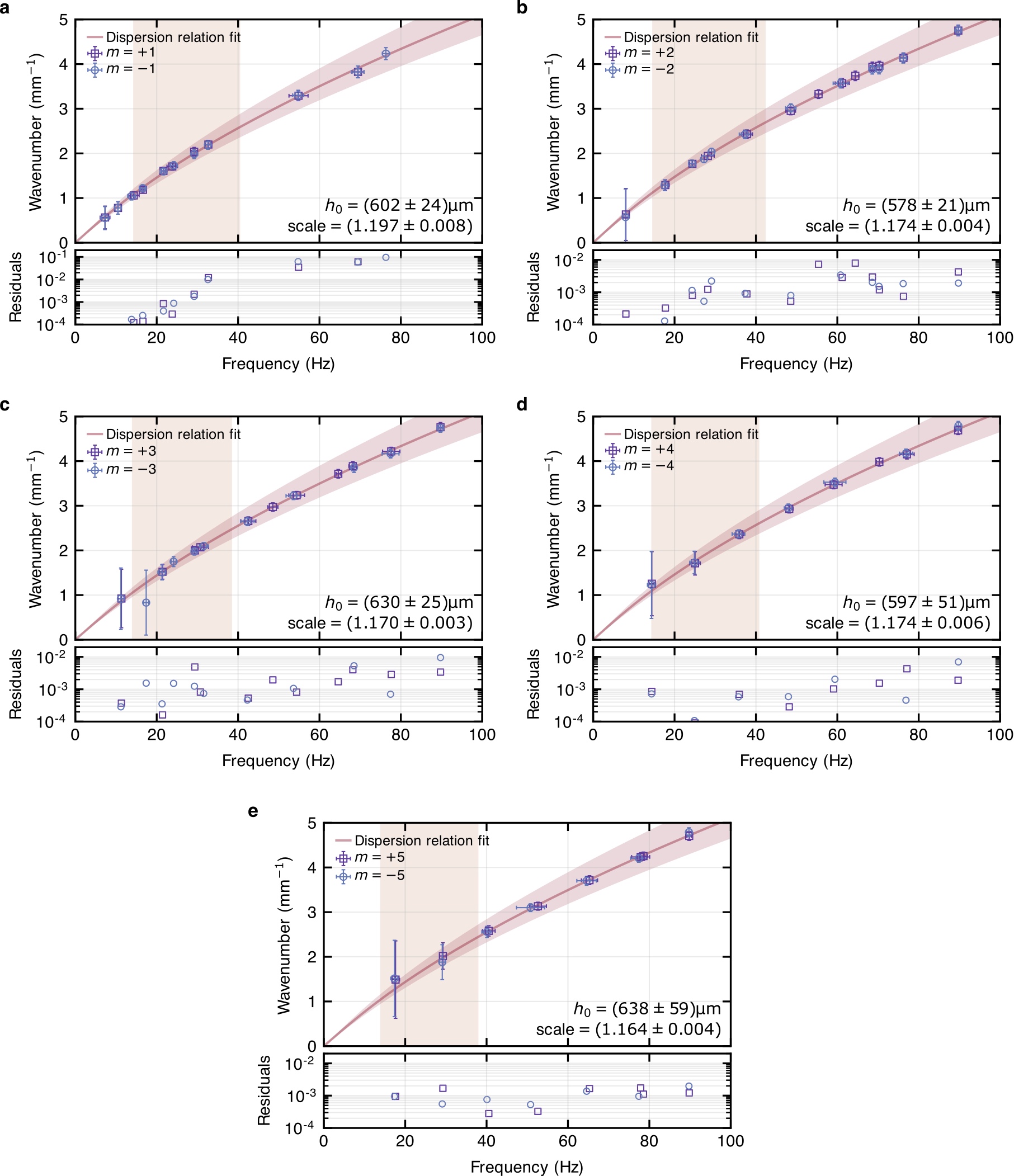}
    \caption{\textbf{Reconstruction of the dispersion relation for different azimuthal modes.} %
    \panel{a-e} Fits of the rescaled dispersion relation (Eq. \eqref{eq:disp}, red line) yield the helium film's thickness, $h_0$ and the scaling factor denoted as scale. Red-shaded regions denote the 1$\sigma$ confidence intervals and the orange-shaded frequency intervals mark the crossover between shallow-water (lower $f$) and deep-water (higher $f$) behaviour. Residuals denote the absolute difference between experimental and theoretical fit of the radial profiles at each frequency, used to determine the wavenumbers.}
    \label{fig:s:disp}
\end{figure*}

From the data set discussed in Sec.~\ref{sec:fridge}, we extract a number of modes with $|m| \in \{1,2,3,4,5\}$; positive (negative) values denote waves that co-rotate (counter-rotate) with respect to the coordinate system. In Fig.~\ref{fig:s:disp}, we carry out five independent reconstructions of the waves' dispersion relation. Consistency of the obtained fitting parameters underlines the robustness of our approach, allowing us to combine these parameters and estimate $h_0 = (602 \pm 13)~\mathrm{\mu m}$.

\section{\label{suppl:vortex}Persistent superflows and wave dissipation due to quantum vortices}

Due to quantum-mechanical effects, vorticity in superfluids is localised on singular structures known as quantum vortices~\cite{barenghi2023book}. In superfluid \textsuperscript{4}He, each quantum vortex induces an irrotational whirl around its core, such that the circulation (the loop integral of flow velocity) equals elementary quantum $\kappa \approx 0.1~\mathrm{mm^2/s}$. In superfluid thin films, quantum vortices may extend from the substrate to the free superfluid interface, collectively inducing a macroscopic, persistent superflow analogous to persistent electrical currents in superconducting coils.

Pairs of surface modes counter-propagating on this background flow experience opposite Doppler shifts, resulting in frequency splitting, whose magnitude generally depends on the radial profile of the mode and the spatial arrangement of quantum vortices. When the vortices are regularly spaced, the induced superflow resembles solid-body rotation with some tangential speed $u$ at the film's outer radius ($R = 5$~mm in our case). The relative frequency splitting can then be expressed as $\Delta f/f = \gamma u/c$, where $\gamma = 0.409$ for the $|m|=2$ Bessel modes discussed in Sec.~\ref{ssec:results-wet} \cite{ellis1993quantum}, and $c = 75$~mm/s is the propagation speed of shallow-water waves on our film (see Sec.~\ref{sec:disc}). To relate the frequency splitting to the number of quantum vortices $N$, we compare the circulation of the vortex-induced superflow, $2\pi R u$, with its discrete equivalent, $N\kappa$. Rearranging yields
\begin{equation}
    N = \frac{2\pi R c}{\gamma\kappa} \left(\frac{\Delta f}{f}\right).
    \label{eq:vort}
\end{equation}
For the employed camera frame rate and image acquisition time, $\Delta f = 0.1$~Hz and the maximum accessible frequency by Fourier transform is $f = 100$~Hz. Thus, we estimate from \eqref{eq:vort} that $N \approx 60$ quantum vortices should induce a principally detectable frequency split in the $|m|=2$ azimuthal band. This estimate increases when considering waves within the shallow-water approximation or when requiring $\Delta f$ to be larger than the frequency resolution.

In practice, quantum vortices can exhibit either positive or negative circulation, with opposing effects on mode frequency splitting. Therefore, $N$ represents the net difference between the populations of positively and negatively oriented vortices, a condition seemingly not achieved in our system. This interpretation aligns with observations in \cite{Sachkou2019-jk}, where the measured splitting arises from spatially separated bundles of vortices with opposite circulation.

Moreover, the presence of quantum vortices in superfluid films provides a means of energy dissipation for surface waves, discussed in the context of relatively thick superfluid films in~\cite{penanen2002model}. This mechanism can explain why spectral peaks presented in Fig.~\ref{fig:results-dry}a acquire a finite width associated with quality factors smaller than $125$, as reported in the main text.

\clearpage
\twocolumngrid
\interlinepenalty=10000
\bibliography{references}
\clearpage\raggedbottom

\end{document}